\documentclass[twocolumn,english,aps,showpacs,prb,superscriptaddress,longbibliography]{revtex4-2}
\usepackage[T1]{fontenc}
\usepackage{color}
\usepackage{bm}
\usepackage{amsmath}
\usepackage{amssymb}
\usepackage{graphicx}
\usepackage{esint}
\usepackage{mathrsfs}
\usepackage{booktabs}
\usepackage{multirow}
\usepackage{makecell}
\usepackage{verbatim}
\usepackage[plainpages = false, pdfpagelabels, 
bookmarks,
bookmarksopen = true,
bookmarksnumbered = true,
breaklinks = true,
linktocpage,
colorlinks = true,
linkcolor = blue,
urlcolor  = blue,
citecolor = blue,
anchorcolor = green,
hyperindex = true,
hyperfigures
]{hyperref} 
\usepackage{dcolumn}   

\makeatletter
\newcommand*{\rom}[1]{\expandafter\@slowromancap\romannumeral #1@}
\makeatother

\usepackage{times}{\Large}

\newcommand{\unit}[1]{\,\mathrm{#1}} 
\newcommand{\equa}[1]{Eq.~\eqref{#1}} 
\newcommand{\fig}[1]{Fig.~\ref{#1}}

\newcommand{\apdx}[1]{Appendix~\ref{#1}}

\renewcommand{\vec}{\mathbf}

\graphicspath{{figures/}}

\begin{document}

\title{Floquet engineering of many-body states by the ponderomotive potential}
	
	\author{Zhiyuan Sun}
	\affiliation{Department of Physics, Tsinghua University, Beijing 100084, P. R. China}
	\affiliation{Frontier Science Center for Quantum Information, Beijing 100084, P. R. China}


	\begin{abstract}
		The ponderomotive force is an effective static  force that a particle feels in an oscillating  field, whose static potential  may be called the ponderomotive potential. We generalize this notion to  periodically driven quantum many-body systems, and propose it as a convenient tool to engineer their non-equilibrium steady states beyond the single particle level. 
		Applied to materials driven by light, the ponderomotive potential is intimately related to the equilibrium optical conductivity, which is enhanced close to resonances. We show that the ponderomotive potential from the incident light may be used to induce exciton condensates in semiconductors, to generate attractive interactions leading to superconductivity in certain electron-phonon systems, and to create additional free energy minima in systems with charge/spin/excitonic orders. These effects are presented with experimentally relevant parameters.		
	\end{abstract}
	
\maketitle

\section{Introduction}

There has been widespread interest in the non-equilibrium phenomena of many-body systems driven by a force that oscillates periodically in time~\cite{
	Sieberer.2016,
	Basov:2017_review,
	Eckardt.2017_rmp_amo,
	Rudner:2020aa,
	Kennes.2021_rmp, 
	Zhou.2022_review,
	Ho:2023aa,
   Murakami.2023_rmp_photoinduced, 
	Sieberer.2023_review, Mori:2023aa}.
 A common example is solid state materials driven by light in  pump probe experiments~\cite{Mahmood:2016aa, Basov:2017_review,  McIver:2020aa, Shan:2021aa, Kennes.2021_rmp, Zhou.2022_review, Zhou.2023_Floquet_black_phosphorus, Murakami.2023_rmp_photoinduced}. This periodic drive can be viewed as a controlling knob that renders the materials in non-equilibrium steady states (NESS) that have properties absent in equilibrium, realizing `Floquet Engineering'~\cite{Lindner:2011aa, Bukov.2015_floquet_review, Moessner:2017ur,
 Oka:2019aa,
 Rudner:2020aa, Kennes.2021_rmp, Zhou.2022_review, Mori:2023aa, Zhou.2023_Floquet_black_phosphorus,Huber.2023}. Floquet engineering of single particle properties has been widely studied and well understood~\cite{Oka.2009, Demler.2010_floquet,	Refael.2015_bath,
Mahmood:2016aa,  He:2019aa, McIver:2020aa, Rudner:2020aa, Yang.2021,  Zhou.2023_Floquet_black_phosphorus,Huber.2023}, while that of systems with many-body interactions has remained problematic~\cite{Bukov.2015_floquet_review,Goldman.2014,Kuwahara.2016, 
Abanin.2017_floquet_manybody, Claassen:2017_spin_liquid_floquet,
Wan.2017_Frustrated_Magnet_floquet, Moessner:2017ur, Sentef.2017_SC_CDW,  Schuler.2020, Kennes.2021_rmp,  Ho:2023aa, Murakami.2023_rmp_photoinduced, Sieberer.2023_review, Wang.2023_LaH10, Berthier.2001}.	

\begin{figure}
	\includegraphics[width= \linewidth]{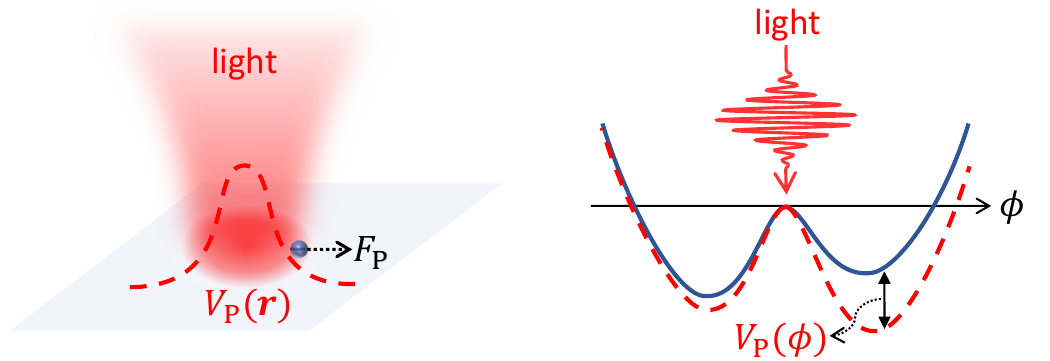} 
	\caption{Left:  An inhomogenenous optical field imposes the conventional ponderomotive potential $V_{\text{P}}(r)$ in real space for a particle. Right: Uniform light imposes a generalized ponderomotive potential $V_{\text{P}}(\phi)$ for a collective degree of freedom $\phi$ in a many-body system. It modifies the energy landscape from the blue curve to the red dashed one.}
	\label{fig:P_potential}
\end{figure}

The ponderomotive force~\cite{Aliev:1992aa,  Sun.2018,Wolff.2019_P_force_graphene, Rikhter.2024} refers to the static second order force $F_{\text{P}}=- \nabla [e E(r)]^2/(4m \omega^2)$ that a particle with charge $e$ and mass $m$ feels  in an inhomogeneous electric field $\vec{E}(r)\cos(\omega t)$ oscillating at frequency $\omega$. 
It originates from the particle's fast out-of-phase oscillation following the electrical force: when the fast force points to the direction of decreasing field, the particle locates closer to the strong field region, and vice versa, leading to a nonzero time average of the net force.
One may define a ponderomotive potential by $F_{\text{P}}=-\nabla V_{\text{P}}(r)$ in the real space coordinate $r$, as shown in \fig{fig:P_potential}.
This notion is also the underlying physics for the Kapitza pendulum~\cite{Kapitza.1951}, and the optical lattices~\cite{ Grimm:2000_optical_trap}  and tweezers~\cite{Moffitt.2008} that  trap atoms and other objects.	
In this paper, we generalize the  ponderomotive potential from $V_{\text{P}}(r)$ to $V_{\text{P}}(\phi)$ where $\phi$ means the generic degrees of freedom in periodically driven  many-body systems, and show that it offers a convenient tool to engineer their NESS, as illustrated in \fig{fig:P_potential}.
\\

%

\section{Ponderomotive Potential}

The Lagrangian of a generic periodically driven system may be written as

\begin{align}
L=L_{\text{f}}[X,\phi] -2 f  \cos(\omega t) \cdot P[X,\phi]  + L_{\text{s}}[\phi]
\,
\label{eqn:action}
\end{align}
where $X$ is the collection of the fast degrees of freedom, $\phi$ is the slow one, $f \cos(\omega t)$ is the periodic driving force with frequency $\omega$, and $P(X,\phi)$ is the generalized polarization the force couples to.  
We assume that the driving term is slowly turned on, and that there is a bath (e.g., the phonons and the substrate) taking away any excessively generated heat so that the system is in a NESS. 
Note that in principle, the system should be described by a Keldysh path integral on the Keldysh time contour~\cite{Kadanoff.1962,Keldysh.1964, Kamenev.book, Altland.2010, Sieberer.2016}. However, the physics could be understood qualitatively with the plain Lagrangian.
For notational simplicity,  we use \equa{eqn:action} to represent the Keldysh action, and keep its full form in the appendices for curious readers.

The central goal of this paper is to integrate out $X$ to obtain an effective Lagrangian or Hamiltonian for the slow field $\phi$ with an energy cutoff $\omega$, whose ground/thermal state is a good approximation to the NESS. 
This effective Lagrangian may be written as 
\begin{align}
L_{\text{P}}[\phi,f]= L_{\text{s}}[\phi] + V_0[\phi] + V_{\text{P}}(\phi,f)
\,,\quad
F_{\text{P}} = -\partial_\phi  V_{\text{P}}
\label{eqn:V_P}
\end{align}
where $V_{\text{P}}$  is  the static `ponderomotive potential' and $F_{\text{P}}$ is the generalized  ponderomotive force for the slow field $\phi$, following the terminology for single particles. 
$V_0$  is the equilibrium term irrelevant to the drive.
Note that the Lagrangian in our convention has a sign difference to the text-book one such that a lower potential means lower Lagrangian.
We expand $V_{\text{P}}$  in even powers of the driving force $f$:
\begin{align}
V_{\text{P}} = \sum_{n=1}^{\infty} \chi^{(2n-1)}(\phi) f^{2n}
\,.
\label{eqn:vp_expansion}
\end{align}
If the driving term has multiple frequencies with components $f_1(\omega_1)$, $f_2(\omega_2)$, ..., one just sums over all their products that combine to zero frequency.
We start with a statement:

The coefficients in \equa{eqn:vp_expansion} are just the real parts of the retarded response functions of $P$ to $f$ at fixed $\phi$ in equilibrium:
\begin{align}
&	\chi^{(1)}(\phi) = - \mathrm{Re}\left[\chi_{\text{R}}(\phi;\omega)\right]
	,\,\notag\\
&	\chi^{(3)}(\phi) = - \mathrm{Re}\left[\chi_{\text{R}}^{(3)}(\phi;\omega,-\omega,\omega)
	\right]
,\, ... \,,
\label{eqn:lemma1}
\end{align}
in either of the following two cases:
\emph{Case 1}: There is no dissipation;
\emph{Case 2}: 
$L_{\text{f}}[X,\phi] =L_{\text{f}}[X]$, and $P$ from \equa{eqn:action} could be separated as $P_1(X) P_2(\phi)$.

Specifically, $\chi_{\text{R}}(\phi; \omega)$ is the linear response function of $P$  to $f$ at frequency $\omega$, and $\chi_{\text{R}}^{(3)}(\phi;\omega,-\omega,\omega)$ is the third order nonlinear response ($P(\omega)$ to $f(\omega)^2f(-\omega)$)~\cite{Boyd.2008}. 
For instance, if $L_{\text{f}}[X]=(-\dot{X}^2+  \omega_0^2 X^2)/2$ is that of a Harmonic oscillator and $P=X$, there is only linear response and $V_{\text{P}} =  f^2/(\omega^2-\omega_0^2)$.
A simple interpretation is that, the oscillator is polarized parallel (anti-parallel) to the fast force $f$ when $\omega<\omega_0$ ($\omega>\omega_0$), giving a negative (positive) coupling energy on average (the actual $V_{\text{P}}$ is one half of this average coupling energy due to partial cancellation from $L_{\text{f}}$). 

Case~1 (absence of dissipation) happens, for example, when the driving frequency is not equal to the intrinsic frequency of an ideal Harmonic oscillator, or  when the frequency of an incident light is below the  gap of an insulator (valid at the linear response level). 
In this case, the retarded response functions in \equa{eqn:lemma1} have no imaginary parts (away from poles), so that they are also equal to the advanced and time-ordered Green's functions~\cite{Altland.2010, Kamenev.book}.
Dissipation may arise from  resonant excitation of the system by the drive, or from a bath~\cite{Refael.2015_bath,  Shirai_2016,Ikeda.2020_NESS, Mitra.2014, Mori:2023aa} such that the spectrum of the system has nonzero line-widths~\cite{Wang.2019_Non_Hermitian}. 
One could explicitly add the bath  to \equa{eqn:action} as a degree of freedom $X_{\text{b}}$ that couples to $X$ and $\phi$.  
After integrating out the bath $X_{\text{b}}$, there are dissipation and fluctuation terms for $X$ and $\phi$ captured by a Keldysh action~\cite{Kadanoff.1962,Keldysh.1964, Kamenev.book, Altland.2010, Sieberer.2016,  Mitra.2005, Mitra.2006}.
\apdx{apdx:proof} contains a proof of \equa{eqn:lemma1} by further integrating out the fast degree of freedom $X$ using the Keldysh path integral. 

For systems not covered by cases~1 and 2, one may compute the ponderomotive potential $V_{\text{P}}$ case by case using Keldysh path integral and Green's functions, which are often simplified by classical approaches when quantum fluctuations are small. Classically, the force exerted  on $\phi$ by the fast degrees of freedom  in \equa{eqn:action} is $F_{\phi}=- \partial_\phi(L-L_{\text{s}})$, and the Ponderomotive force  would simple be its time average: $F_{\text{P}}=\langle F_{\phi} \rangle_t$ (keeping only the terms dependent on the driving field $f$). 
In the absence of dissipation from a bath or from heat generation such that $X(t)$ is periodic in time and satisfies the Euler-Lagrange equation of motion,  it is straightforward to show that the ponderomotive potential is simply the time averaged Lagrangian: $V_{\text{P}}=\langle L-L_{\text{s}} \rangle_t$.
To compute its quantum mechanical version, one just replaces the time average by the path integral average: $F_{\text{P}}=\langle F_{\phi} \rangle_{\text{path integral}}$.

Compared to the high frequency Magnus expansion~\cite{Goldman.2014, Bukov.2015_floquet_review} which generates the effective static Hamiltonian by an expansion in the inverse driving frequency $1/\omega$, the ponderomotive potential  in \equa{eqn:vp_expansion} is different in that it is an expansion in the driving field instead, which may be viewed as a re-summation of the Magnus series.
It does not require $\omega$ to be higher than all the energy scales of the system, such that \equa{eqn:vp_expansion} could capture the physics of dissipation~\cite{Ikeda.2020_NESS} and resonances~\cite{Bukov.2016, Murakami.2023_rmp_photoinduced}. The ponderomotive potential may also be obtained from a Schrieffer-Wolff transformation~\cite{Bukov.2016} applied to the Lindblad master equation formulation of a driven dissipative system (represented by \equa{eqn:action}, or more rigorously, \equa{eqnSI:action})  that eliminates the periodic driving term order by order. However, the path integral approach applied here is convenient in separating the slow and fast fields and in making connections to response functions.


\subsection{Materials driven by light}
The most apparent real world application of \equa{eqn:vp_expansion} is  to materials driven by the dynamical electric field $2E \cos(\omega t)$ of light, which happens in, e.g.,  pump-probe experiments~\cite{
	Basov:2017_review,
	Kennes.2021_rmp, 
	Zhou.2022_review}. 
In this case, the response functions (polarizabilities) are simply related to the  linear and nonlinear optical conductivities~\cite{Sun.2018}. For example, the lowest order term in \equa{eqn:vp_expansion} reads
\begin{align}
	V_{\text{P}} = \mathrm{Re}
	\left[-\frac{i}{\omega} \sigma(\phi,\omega)
	\right] 
	E^2
	= \frac{1}{\omega} \sigma_2(\phi,\omega) E^2
	\,
\label{eqn:vp_optical}
\end{align}
where $\sigma(\omega)$ is the optical conductivity of the system for fixed $\phi$ and $\sigma_2$ is its imaginary part. 
Similarly, the $n=2$ term in \equa{eqn:vp_expansion} would be $\mathrm{Re}
\left[-i \sigma^{(3)}(\phi; \omega, -\omega, \omega)/\omega
\right] E^4$ where $\sigma^{(3)}$ is the third order optical conductivity~\cite{Boyd.2008,Sun2018b}, familiar to the nonlinear optics community. 
Since there are often peaks in $\sigma(\omega)$ from resonant transitions, $V_{\text{P}}$ will be resonantly enhanced when the driving frequency is tuned close to these transitions.

If $\phi$ is the coordinate $r$ of a single particle with charge $e$ and mass $m$ moving in an inhomogenenous optical field represented by $E(r)\cos(\omega t)= E_0 g(r) \cos(\omega t)$, the optical conductivity is just $\sigma= g(r)^2 ie^2/( m \omega) $, which plugged into \equa{eqn:vp_optical} and \equa{eqn:V_P} gives the static  ponderomotive force $F_{\text{P}}=- \nabla [e E(r)]^2/(4m \omega^2)$ it experiences~\cite{Aliev:1992aa, Sun.2018}.
If the particle is an atom with an internal optical transition of energy $\omega_0$ and linewidth $\gamma$, and $X$ represents this internal degree of freedom, the ponderomotive potential from \equa{eqn:vp_optical} is then $V_{\text{P}} \propto E(r)^2 \mathrm{Re}[1/(\omega^2-\omega_0^2+ i \gamma \omega)]$. This is the potential that traps atoms in optical lattices and  tweezers~\cite{ Grimm:2000_optical_trap, Moffitt.2008}.

In sections~\ref{sec:exciton condensate}, \ref{sec:sc} and \ref{sec:new_minima}, we apply the ponderomotive potential to three examples to show that one may use it to engineer correlated states in materials driven by light.
\\

\section{Light induced exciton condensate}
\label{sec:exciton condensate}
\begin{figure}
	\includegraphics[width= \linewidth]{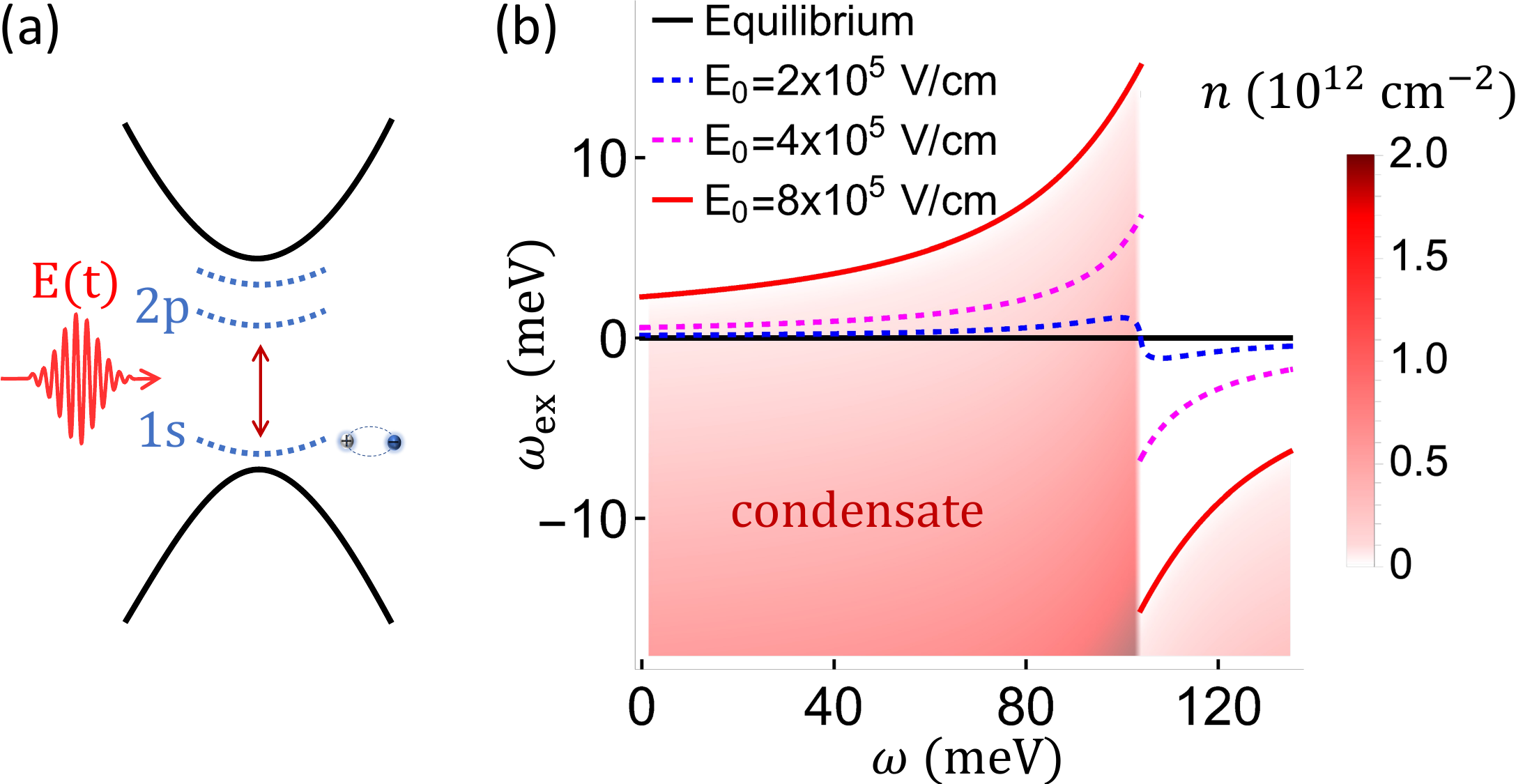} 
	\caption{(a) Schematics of the electron (black solid curves) and exciton (blue dashed curves) dispersion of a semiconductor and the pump light (red curve). (b) The phase diagram of the driven semiconductor on the plane of light frequency $\omega$ and the bare exciton energy. The curves are the boundaries between the semiconductor and condensate phases at different pumping electric fields $E_0$. The red color scale shows the exciton density of the condensate phase for  $E_0= 8 \times 10^5 \unit{V/cm}$. 
		The electron/hole mass is chosen as the vacuum electron mass $m=m_{\text{e}}$. The dielectric is $\epsilon=7.0$, giving a Bohr radius of $a=0.74 \unit{nm}$ and the $1s \rightarrow 2p$ transition energy $\omega_0=104 \unit{meV}$. The damping rate is $\gamma=8 \unit{meV}$.
	}
	\label{fig:exciton_phase_diagram}
\end{figure}

The first example is a semiconductor driven by coherent light at a sub-gap frequency $\omega$.
The exciton spectrum in the semiconductor is shown schematically in \fig{fig:exciton_phase_diagram}(a).  Neglecting spatial fluctuations, the Lagrangian density for the lowest two excitons can be written as~\cite{Sun.2023_dynamical}
\begin{align}
L =&   \Phi_{\text{s}}^\ast 
\left( -i\partial_t+ \omega_{\text{ex}}  
\right)  \Phi_{\text{s}}
+
\Phi_{\text{p}}^\ast
\left( -i\partial_t+ \omega_{\text{p}}  
\right)  \Phi_{\text{p}} + g \rho^2
\notag \\
& 
+ \lambda E(t) (\Phi_{\text{p}}^\ast \Phi_{\text{s}} + c.c.) 
\,
\label{eqn:exciton_Lagrangian}
\end{align}
where $\Phi_{\text{s}}, \omega_{\text{ex}}$ and $\Phi_{\text{p}}, \omega_{\text{p}}$ are the bosonic fields and energies of the $1s$, $2p$ excitons, and $g$ is the strength  for the local interaction between the total density $\rho=\Phi_{\text{s}}^\ast \Phi_{\text{s}} + \Phi_{\text{p}}^\ast \Phi_{\text{p}}$. The dynamical electric field $E(t)=E_0 \cos (\omega t)$ inter converts the two excitons with the matrix element $\lambda=c_{\text{p}} e a$, where $c_{\text{p}} \sim 1$  and $a=\epsilon 2\hbar^2/(m e^2)$ is the Bohr radius  for the electron and hole with mass $m$ and charge $\pm e$ bounded by Coulomb attraction screened by the dielectric $\epsilon$. This coupling leads to the familiar $1s \rightarrow 2p$ optical transition in a Rydberg series. If the electric field is along the $x$ direction, the $\Phi_{\text{p}}$  refers to the $p_{\text{x}}$ exciton among the degenerate $2p$ excitons.

Without the driving term and for $\omega_{\text{ex}}>0$, there are no excitons in the ground state. However, if the binding energy is large enough such that $\omega_{\text{ex}}<0$, the s-excitons spontaneously emerge and form an excitonic superfluid at low temperatures~\cite{Mott1961,Fogler2014a, Sun.2023_dynamical}, as shown by the equilibrium phase boundary (black line) for the semiconductor-to-excitonic insulator transition  in \fig{fig:exciton_phase_diagram}(b). For simplicity, we focus on the zero temperature case and the superfluid density is $\rho_0=-\omega_{\text{ex}}/(2g)$ at the mean field level. 

With the coherent light,  we focus on its effect on the critical regime where  $\omega_{\text{ex}}$ is close to zero, so that $\Phi_{\text{s}}$ is a slow field and $\Phi_{\text{p}}$ is the fast field to be integrated out. 
To be precise, the fast degree of freedom is the internal degree of freedom of an exciton responsible for the 1s to 2p transition, which we use $\Phi_{\text{p}}$ to represent for notational simplicity.
This system satisfies case~2 of \equa{eqn:lemma1} whether there is damping or not. Therefore, the $O(E^2)$ ponderomotive potential for the slow field is obtained from  \equa{eqn:vp_optical}  as
\begin{align}
	V_{\text{P}}(\Phi_s) = \lambda^2 E_0^2 \text{Re}
	\left[
	\frac{\omega_0}{\omega^2-\omega_0^2+ i \gamma \omega} 
	\right]
	 \Phi_s^\ast  \Phi_s
	\,
	\label{eqn:vp_exciton}
\end{align}
where $\omega_0=\omega_{\text{p}}-\omega_{\text{ex}}$ is the transition energy and we have added a damping rate $\gamma$ for the excitons.
$V_{\text{P}}(\Phi_s)$ shifts down/up the effective energy of the 1s-exciton if the driving frequency is red/blue tuned relative to the $1s \rightarrow 2p$ transition, whose effect is resonantly enhanced when $\omega$ is close to $\omega_0$. 
Physically, compared to the optically silent state with no excitons, the state with a nonzero density of excitons has optical transitions with the optical conductivity $\sigma_2$ being negative at a red tuned frequency, giving a negative driving energy in \equa{eqn:vp_optical} which  favors such a state.
As a result, the phase boundary of the nonequilibrium steady state is modified to the blue curve in \fig{fig:exciton_phase_diagram}(b). 
If $\omega_{\text{ex}}$ is just a little above zero, driving the system at a red tuned frequency obviously shifts the system from the semiconductor into the condensate phase, meaning `light induced exciton condensate'.

For stronger driving fields such that $\lambda E_0 \gtrsim |\omega-\omega_0|, \gamma$, the linear response becomes a poor approximation and one needs to sum all the terms in \equa{eqn:vp_expansion}. Physically, the exciton undergoes  oscillation between the $1s$ and $2p$ state where $\Phi_s$ has a fast component too. In this case, we write the excitonic field as $(\Phi_s, \Phi_p)= \Phi \xi$
so that $\Phi =\sqrt \rho e^{i \theta}$ is the total excitonic field, and the $1s$, $2p$ degree of freedom is viewed as the internal degree of freedom of the exciton encoded in the spinor $\xi=(\sqrt{1-\eta}, \sqrt{\eta}e^{i \varphi} )$. The Lagrangian in these variables becomes
\begin{align}
	L &=   \Phi^\ast
	\left( -i\partial_t+ \omega_{\text{ex}}  
	\right)  \Phi
	+ g \rho^2 + \rho L_\xi
	\,,\notag \\
L_\xi &=
	\eta (\dot{\varphi} + \omega_0)
	+ \lambda E(t) \sqrt{\eta(1-\eta)} 2 \cos \varphi
	\,.
	\label{eqn:exciton_Lagrangian_2}
\end{align}
$L_\xi$ is just the Lagrangian of a pseudo spin $\vec{n}=\xi \vec{\sigma} \xi^{\dagger}$ in the pseudo magnetic field $\vec{B}=(\lambda E(t), 0, \omega_0/2)$ where $\vec{\sigma}=(\sigma_{\text{x}}, \sigma_{\text{y}}, \sigma_{\text{z}})$ are the usual Pauli matrices. 

Now the total exciton field $\Phi $ is the slow variable while the pseudo spin is the fast degree of freedom undergoing Rabi oscillation. 
If there is dissipation (formally introduced by the Keldysh action version of it), \equa{eqn:exciton_Lagrangian_2} is not covered by cases~1 and 2. Nevertheless, when there is a condensate such that the spin becomes classical, the Pondermotive force for $\Phi$ can be  computed simply from  $F_{\text{P}}=  \Phi \langle -\partial_\rho (\rho  L_\xi) \rangle_t$ where the time average is on the classical dynamical orbit of the spin. 
The resulting ponderomotive potential  predicts the pink  and red  lines as the phase boundaries for two strong driving fields  in \fig{fig:exciton_phase_diagram}(b). A prominent feature is the discontinuity of the boundary as $\omega$ is tuned across the  resonance, a consequence of the change of the stable orbit of the classical pseudo spin.
See \apdx{apdx:condensate} for details of the 
calculation.

Contrary to previously discussed mechanisms where the excitons are \emph{generated} by light~\cite{Perfetto.2020_pumped_p_exciton, Butov1994}, this phenomenon doesn't require any interband optical matrix elements, but just the Hydrogenic $s \rightarrow p$ optical matrix element of an exciton, which  naturally exists. 
As one slowly turns on an optical field  in a semiconductor which does not have any optical transitions, the drive gradually lead the system into an exciton condensate (which finally have optical transitions). Of course, the electrons and holes in the excitons must come from somewhere in this process, which could be the electrical contacts or  small  interband matrix elements in realistic devices.
Therefore, this effect works for generic devices including those with interlayer excitons~\cite{Butov1994, Ma.2021strongly, Wang.2023_InAs_GaSb}. 
In devices made of GaAs~\cite{Butov1994} and transition metal dichalcogenides~\cite{Ma.2021strongly}, the lowest exciton energy is at the order of  $\omega_{\text{ex}} \sim 1 \unit{eV}$.
To tune $\omega_{\text{ex}}$ to the energy range in \fig{fig:exciton_phase_diagram}, a practical method in the near term is to apply an electrical contact bias $\mu$ (to be distinguished from a huge gating field potentially causing dielectric breakdown) that shifts down the effective exciton energy to $\omega_{\text{ex}}-\mu$ ~\cite{Ma.2021strongly, Sun.2023_dynamical}.

This phenomenon may be partly understood as the optical Stark effect~\cite{Frohlich.1985} where the  $s$-exciton energy is pushed down in the Floquet picture for a red tuned driving frequency. However, the effective action approach in \equa{eqn:V_P} is indispensable  in determining the many-body steady state, especially when there is dissipation and a condensate.
\\

\section{Light induced superconductivity}
\label{sec:sc}

The second example is light induced superconductivity in an electron-phonon system in a metal. It is described by the  Lagrangian density
\begin{align}
L =&  \frac{1}{2} 
\left[- \dot{X}^2 + (\omega_0^2 + g_{\text{e}} \rho ) X^2  
\right]
+ E(t) X + L_{\text{e}}
	\,
	\label{eqn:electron_phonon}
\end{align}
where $L_{\text{e}}$ is the electronic Lagrangian, and $X$ represents an infrared (IR) active phonon which couples linearly to the pump electric field $E(t)=2E_0 \cos (\omega t)$ but nonlinearly to the local electron density fluctuation $\rho(r)$ due to inversion symmetry~\cite{Kennes2017,Sentef.2017_IR_phonon_attraction, Grankin.2021_IR_phonon_attraction, Michael.2023_light_induced_SC, Kovac.2023_light_induced_SC, Yarmohammadi2023}. 

\begin{figure}
	\includegraphics[width= 0.9 \linewidth]{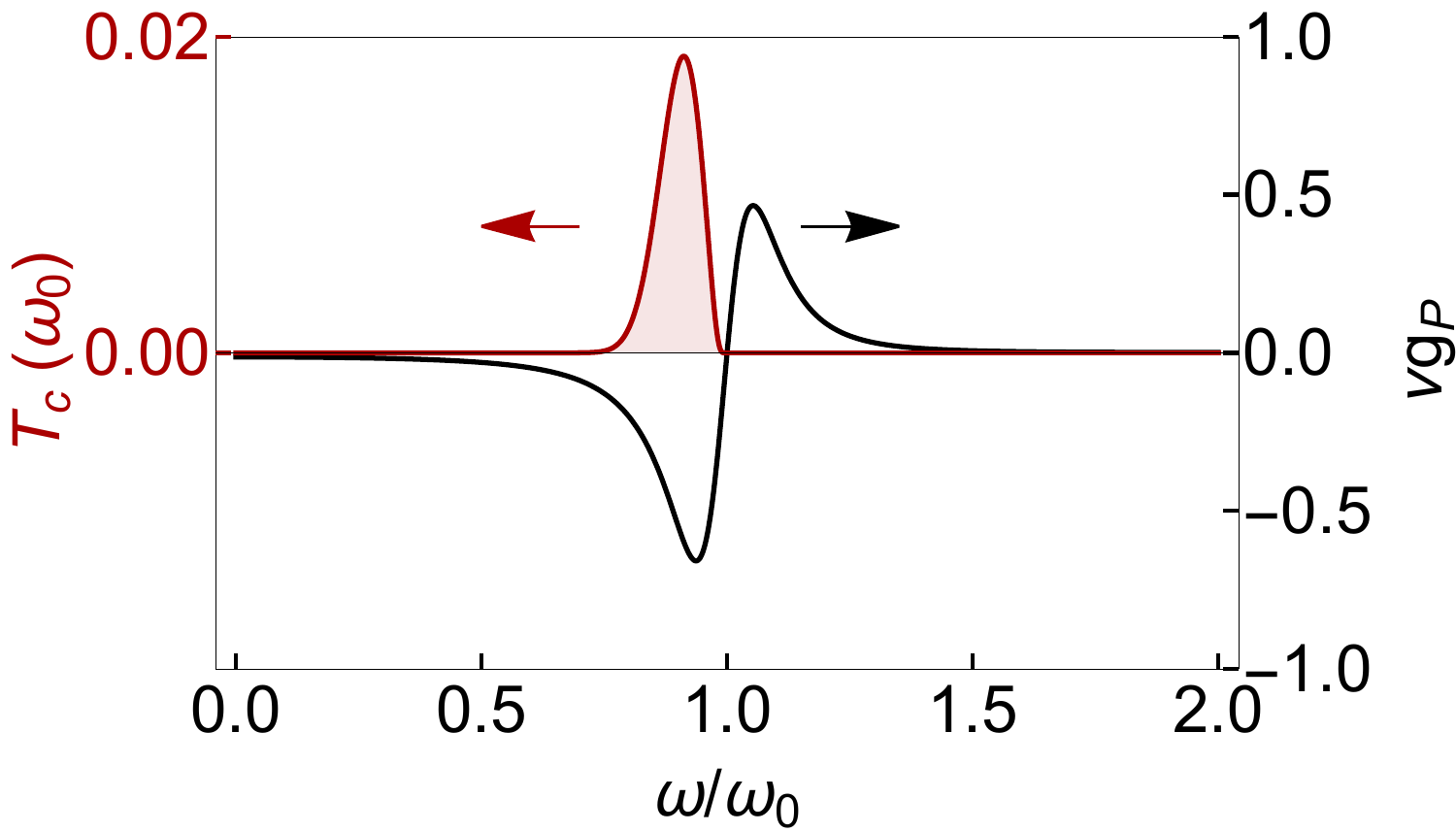} 
	\caption{The black curve is the dimensionless  electron-electron interaction $\nu g_{\text{P}}$ from \equa{eqn:vp_ep_damping} as a function of the light frequency $\omega$ for fixed driving electric field $E_0$. 
		The red curve is the estimated	superconducting $T_{\text{c}}$ in units of $\omega_0$.
		Plotted is for an infrared active phonon with the intrinsic frequency $\omega_0=10 \unit{THz}$ and the damping rate $\gamma=2 \unit{THz}$ and other parameters specified in appendix~\ref{apdx:k3c60}.}
	\label{fig:superconductivity_phase_diagram}
\end{figure}

Treating low energy electrons as the slow field and $X$ as the fast field, the
ponderomotive potential for the electron density is found from \equa{eqn:vp_expansion} as
\begin{align}
	V_{\text{P}}(\rho) =  \frac{E_0^2}{\omega^2-(\omega_0^2+ g_{\text{e}} \rho)}  
	\xrightarrow{O(\rho^2)}
	\frac{g_{\text{e}} ^2 E_0^2}{(\omega^2-\omega_0^2)^3} \rho^2
	\,.
	\label{eqn:vp_ep}
\end{align}
Note that there are no higher order terms in $E_0$ or quantum mechanical effects  since a Harmonic oscillator has no nonlinear responses.  
The generated local density-density interaction  is thus attractive/repulsive when $\omega$ is red/blue tuned relative to the phonon frequency $\omega_0$, and experiences resonant enhancement when $\omega \approx \omega_0$, as shown by the black curve in \fig{fig:superconductivity_phase_diagram}. This explains previous results by Kennes et al.~\cite{Kennes2017} and Sentef~\cite{Sentef.2017_IR_phonon_attraction} and provides further insights.

When there is dissipation such that the phonon has a damping rate $\gamma$, the situation is not covered by case~1 or 2. By integrating out $X$ in the Keldysh path integral (see \apdx{apdx:SC}), the ponderomotive potential is obtained as
\begin{align}
	V_{\text{P}} &= 
	\frac{E_0^2}{\gamma \omega}
\arctan
	 \frac{\gamma \omega}{\omega^2-(\omega_0^2+ g_{\text{e}} \rho)}  
 \xrightarrow{O(\rho^2)}
 g_{\text{P}} \rho^2
 \,,
\notag \\
g_{\text{P}} &=	\frac{ \omega^2-\omega_0^2}{\left[(\omega^2-\omega_0^2)^2+\gamma^2 \omega^2 \right]^2}
g_{\text{e}}^2 	E_0^2 
	\,.
\label{eqn:vp_ep_damping}
\end{align}
Therefore, damping leads to broadening of this effect as shown in \fig{fig:superconductivity_phase_diagram}. Note that the branch of `arctan' should be selected such that $V_{\text{P}}$ is continuous across the resonance.

Thanks to the exact quantum-classical correspondence  in the  response property of a Harmonic oscillator, \equa{eqn:vp_ep_damping} can also be derived from  the classical equation of motion. The force for the electron density is just the square of the phonon displacement: $F_\rho =  -\partial_\rho (L-L_{\text{e}})  =   -g_{\text{e}} X^2/2 $.  Its time average gives the ponderomotive force
$
F_{\text{P}}
=-g_{\text{e}} |\chi_{\text{R}}(\omega)|^2  E_0^2
\,
\label{eqn:fp_ep_damping}
$
whose integral  over $\rho$ gives \equa{eqn:vp_ep_damping}.
Therefore, $F_{\text{P}}$  pushes $\rho$ in the direction that reduces the phonon frequency, so that the `driving energy' $V_{\text{P}}$ (\equa{eqn:vp_ep})  for a Harmonic oscillator goes down. Across the resonance, $F_{\text{P}}$ has a peak rather than changing its sign as expected naively.

For a spherical fermi surface, the dominant instability due to the  attraction is towards the BCS superconductivity~\cite{Shankar.1994}.
 The resulting superconducting transition temperature $T_{\text{c}} \approx 1.13 \Lambda e^{1/(\nu g_{\text{P}})}$ for $g_{\text{P}}<0$ is shown in \fig{fig:superconductivity_phase_diagram}, where the electronic density of states $\nu$ in the normal state and the other parameters are estimated in \apdx{apdx:k3c60} for  the fulleride K$_3$C$_{60}$~\cite{Kennes2017}, a possible system to test \equa{eqn:vp_ep_damping}.
The energy cutoff is chosen as $\Lambda=|\omega_0-\omega|$ because $g_{\text{P}}$ depends the frequency $\omega_{\text{e}}$ of  $\rho$ (to be distinguished from $\omega$ or $\omega_0$) if one goes beyond the static $\rho$ approximation in \equa{eqn:vp_ep_damping}, whose  behavior is expected to change  for $\omega_{\text{e}} \gtrsim |\omega_0-\omega|$~\cite{Gao2020}. 
A detailed `strong coupling' calculation for $T_{\text{c}}$ taking into account the $\omega_{\text{e}}$ dependence  deserves future study. 

To verify this effect in ultrafast experiments, one may use a multi-cycle pump pulse with a bandwidth smaller than the linewidth of the IR phonon, and measure the transient state (similar to recent experiments~\cite{Rowe.2023_k3c60, Michele.2021} but with relatively well defined pump frequencies close to the relevant IR phonon,  and with a probe pulse overlaping with the pump instead of after the pump is gone).  An interesting prediction of \fig{fig:superconductivity_phase_diagram} is that superconductivity is enhanced/suppressed if the pump frequency $\omega$  is lower/higher than the IR phonon frequency $\omega_0$. 
If the equilibrium state is already superconducting, the pump would lead to a Fano-like asymmetric lineshape of $T_{\text{c}}$ as $\omega$ is scanned across $\omega_0$.
Possible systems include K$_3$C$_{60}$~\cite{Rowe.2023_k3c60} and doped SrTiO$_3$~\cite{Marel2019}.
See \apdx{apdx:heating} for estimates of heating. 


A different scenario is the state after a short pump pulse,  when  the IR phonon oscillates freely with an amplitude $X_0$~\cite{Kennes2017, Rowe.2023_k3c60}. If the electron density varies slowly, the oscillation amplitude changes adiabatically with the action $S \propto \sqrt{\omega_0^2 + g_{\text{e}} \rho} X_0(\rho)^2$ (the adiabatic invariant) kept constant. 
Therefore, the ponderomotive force is $F_{\text{P}}(\rho)= -\langle g_{\text{e}} X^2/2 \rangle_t =- g_{\text{e}} X_0(\rho)^2/4= -
 \frac{1}{4} X_0(0)^2 |\omega_0|/\sqrt{\omega_0^2 + g_{\text{e}} \rho}$, rendering the potential $V_{\text{P}}=\frac{1}{2}  |\omega_0| \sqrt{\omega_0^2 + g_{\text{e}} \rho} X_0(0)^2$, generalizing the result of  Kennes et al.~\cite{Kennes2017}. At order $\rho^2$, the density-density interaction strength $g_{\text{P}}=-X_0(0)^2/\left( 16 \omega_0^2 \right)$  is always negative. Since the excitation efficiency of the IR phonon by the pump has a peak at its resonance, it is consistent with a recent experiment on K$_3$C$_{60}$~\cite{Rowe.2023_k3c60}. Nevertheless, this explanation serves only as a possibility of what happened in K$_3$C$_{60}$~\cite{Mitrano2016, Cantaluppi2018, Budden2021, Rowe.2023_k3c60}, among other mechanisms such as the one involving  Jahn-Teller Hg phonons~\cite{Chattopadhyay2023}. The actual mechanism for the light induced superconducting-like state requires more involved study, especially considering its meta-stability~\cite{sun2019transient,Chattopadhyay2023}.

The pondermotive potential also provides a simple derivation for the light enhanced electron-electron attraction in a Raman phonon model~\cite{Babadi.2017, Murakami.2017_ligh_induced_SC}, see~\apdx{apdx:Raman_phonon}. 
\\

\section{Light engineered free energy landscape}
\label{sec:new_minima}

The third example is light induced  new free energy minima in systems with excitonic order (excitonic insulator~\cite{Jerome1967}) or charge/spin density wave (CDW/SDW) order~\cite{Gruner.1988_RMP_CDW, Gruner.1988_RMP_SDW}  due to fermi surface nesting in the BCS weak coupling case. The Lagrangian density~\cite{sun2020topological, sun.2020_BaSh} is 
\begin{align}
L = 
	\psi^\dagger 
\begin{pmatrix}
-i\partial_t  + \xi(p) & \Delta \\
\Delta^\ast   &  -i\partial_t - \xi(p)
\end{pmatrix}
	\psi
+ 
\frac{1}{g}|\Delta|^2 
\label{eqn:BCS_HS_action}
\end{align}
where $p=-i \nabla+A(t)$, $A(t)=2A_0 \cos \omega t$ is the vector potential of the coherent light, $g$ is the coupling constant, and $\psi=(\psi_1, \psi_2)^T$ is the two-component fermion field for two electronic bands with energy $\xi(p)/-\xi(p)$. They correspond to the overlapping conduction and valence bands ($\xi(p)=p^2/(2m) - \mu$) in the case of excitonic order, and the left and right moving bands ($\xi(p)=v_\text{F} p$) nested together by twice the fermi wave vector  in the case of charge/spin order where $v_F$ is the Fermi velocity. The complex order parameter $\Delta$  leads to a quasi-particle gap.  

We treat $\Delta$ as the slow field and  $\psi$ as the fast one.
If $\omega<2\Delta$ and there is no bath such that there is no dissipation, \equa{eqn:BCS_HS_action}  satisfies case~1 of \equa{eqn:lemma1} so that \equa{eqn:vp_optical} gives the ponderomotive potential
\begin{align}
V_{\text{P}}(\Delta) 
	=  \frac{E_0^2}{\omega^2} \frac{n}{m}  \left(1
	-  \frac{2\Delta}{\omega} \frac{\mathrm{sin}^{-1}\left(\frac{\omega}{2\Delta}\right)}{\sqrt{1-\left(\frac{\omega}{2\Delta}\right)^2}}
	 \right) 
	\,
	\label{eqn:vp_BCS}
\end{align}
at order $E_0^2$, where $E_0=\omega A_0$ is the amplitude of the driving electric field and $n$, $m$ are the density of carriers  in the normal state and their effective mass, and we have set the elementary charge to be unity.
The optical conductivity $\sigma$ at zero temperature is taken from Eq.~12 of Ref.~\cite{sun.2020_BaSh} neglecting the BaSh mode. 
Since $V_{\text{P}}$ is negative for $2\Delta>\omega$ and diverges as $-1/\sqrt{\delta}$ as the detuning $\delta=2\Delta-\omega$ approaches zero (a result of the interband transitions at $\omega>2\Delta$), it tends to push $\Delta$ to smaller values.  

Note that for $U(1)$ invariant CDW/SDW systems, unlike excitonic insulators, the gapless phase mode shifts the optical absorption from quasiparticle excitations to zero frequency~\cite{Gruner.1988_RMP_CDW, Gruner.1988_RMP_SDW}. However, this effect is absent in most materials due to strong pinning by the lattice and disorder~\cite{Gruner.1988_RMP_CDW, Gruner.1988_RMP_SDW}, such that the optical conductivity used in \equa{eqn:vp_BCS} still holds.

\begin{figure}
	\includegraphics[width= 0.95 \linewidth]{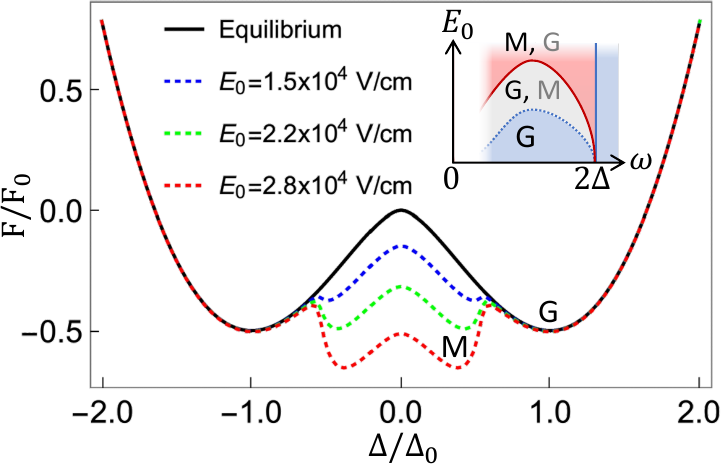} 
	\caption{The effective free energy  for a system with  charge/spin/excitonic order driven by light at several field strengthens $E_0$ for $\omega=30 \unit{THz}$ and $\gamma=1 \unit{THz}$. 
		The gap at equilibrium is $2\Delta_0=55 \unit{THz}$, at the same order as that in Ta$_2$NiSe$_5$~\cite{Lu2017}.
		The order parameter $\Delta$ and the free energy $F$ are in units of  $\Delta_0$ and  $F_0=\nu \Delta_0^2$.
		The top right inset is a schematic phase diagram on the plane of driving frequency $\omega$ and field strength $E_0$. Black symbol means the global minimum while gray symbol means  a metastable minimum. The white region is beyond the validity of the analytical formula in this work.
	}
	\label{fig:landscape}
\end{figure}

To study the case of $\omega > 2\Delta$ when there is absorption, one  needs a bath that takes away the heat, which is implicitly treated as a damping rate $\gamma$ of the quasi particles. The system is  no longer covered by cases~1 or 2, and one must compute $V_{\text{P}}$ by explicitly integrating out $\psi$ in the Keldysh action, giving a lengthy expression for the ponderomotive force $F_{\text{P}}(\Delta)$, as in \apdx{apdx:CSE}. 
It corrects the effective free energy to
\begin{align}
F= - \nu |\Delta|^2 \ln \frac{\Lambda}{|\Delta|}
+ \frac{1}{g} |\Delta|^2 	
+ V_{\text{P}}(\Delta)
\,
\label{eqn:F_BCS}
\end{align}
at zero temperature where $\Lambda$ is the energy cutoff.

The landscape $F(\Delta)$ is plotted in \fig{fig:landscape} for several driving electric fields.  Being negative, $F_{\text{P}}$ always pushes $\Delta$ to smaller values. Notably,  the quasi-particle excitation peak at $\Delta \lesssim \omega/2$ contributes a `dissipative peak' in $F_{\text{P}}$ that scales inversely with $\gamma$. 
Physically, each Anderson pseudo spin $\sigma$ (in the diagonalized band basis) close to the gap edge contributes a force  $\propto \sigma_z$ for $\Delta$. Resonant  optical excitation reduces $\sigma_z$ by an amount $\propto E_0^2/\gamma^2$,  contributing a negative ponderomotive force on top of the equilibrium force.
This force leads to the abrupt drop of the potential around $\Delta=\omega/2$ in \fig{fig:landscape}, which  creates a new minimum labeled by `M' for strong enough fields. As the light intensity grows further, the energy of this minimum goes below that of the original minimum `G', and the system undergoes a first order phase transition from G to M, as shown by the red solid curve in the schematic non-equilibrium  phase diagram in the inset of \fig{fig:landscape}. 
In an ultrafast experiment, one could measure the quasi-particle gap by a probe pulse overlapping with a multi-cycle pump pulse. The non-equilibrium  phase transition would manifest as a sharp drop of the gap as the pump fluence is  increased beyond the threshold.
\\


\section{Discussion}
%

We have introduced the ponderomotive potential by Eqs.~\eqref{eqn:action}\eqref{eqn:V_P}\eqref{eqn:vp_expansion} 
to study periodically driven many-body systems. 
We proved its intimate relation to the equilibrium response functions in
Eq.~\eqref{eqn:lemma1}, which in the context of materials driven by light are shown to be the familiar linear (Eq.~\eqref{eqn:vp_optical}) and nonlinear optical conductivities, offering a quick and convenient way to obtain physical insights.
We also applied it to three realistic examples and found interesting light induced states to be verified by experiments. 
With this concept, we anticipate more exotic non-equilibrium steady states to be discovered, 
especially in systems with a manifold of (nearly) degenerate low energy states~\cite{Sun.2023_dynamical, Sentef.2017_SC_CDW, Wan.2017_Frustrated_Magnet_floquet} such that even a weak drive could seamlessly engineer the energy landscape. 

Although we assumed a bath that takes away the heat such that the NESS exists, the ponderomotive potential may also apply to the quasi steady states in the prethermal stage of driven isolated systems~\cite{Kuwahara.2016, Abanin.2017_floquet_manybody, Goldman.2014, Ho:2023aa}.
In conclusion, we expect the ponderomotive potential to be a convenient tool for studying the non-equilibrium phenomena in  driven systems. To go beyond it, one may look for non-conservative ponderomotive  forces, or could perform systematic expansions to include the  Kinetic terms in the effective low energy Lagrangian.
\\


\begin{acknowledgements}
This work is supported by National Natural Science Foundation of China (Grant No. 12374291) and the startup grant from Tsinghua University.
Z. S.   is grateful to M. M. Fogler, A. J. Millis, Y. Murakami, D. Gole{\v{z}}, T. Kaneko, S. Zhou, Q. Yang, S. Xu, S. Ostermann and C. Huang for helpful discussions.
\end{acknowledgements}

\appendix

\begin{widetext}

\section{Proof of \equa{eqn:lemma1}}
\label{apdx:proof}
It may look natural that the coefficients in \equa{eqn:vp_expansion}
are roughly the response functions  if one `integrates out' the fast degree of freedom $X$ assuming fixed $\phi$.
However, since one is dealing with the non-equilibrium case of periodically driven systems, this integrating out procedure has to be done by the real time path integral, for which a natural framework is the path integral on the Keldysh time contour~\cite{Kadanoff.1962, Keldysh.1964, Kamenev.book, Altland.2010, Sieberer.2016}. 
In the Keldysh path integral~\cite{Kamenev.book,Altland.2010}, one defines a generating functional, the `partition function', as
\begin{align}
	Z[f(s)] &=\mathrm{Tr}[\hat{U}(s) \hat{\rho}]
		=
	\int_C D[X(s), \phi(s)]
	e^{-i S[X, \phi, f]}
		=
\int D[X_c, X_q; \phi_c, \phi_q]
e^{-i S[X_c, X_q; \phi_c, \phi_q; f]}
	\,
	\label{eqnSI:Z}
\end{align}
on the closed time contour $C$ parameterized by $s$, which runs from time $t=0$ to time $t=t_f$ (forward contour) and then from time $t=t_f$ to time $t=0$ (backward contour). 
Here $\hat{\rho}$ is the initial density matrix of the system at time zero, $X_{c/q}(t)=(X_{+} \pm X_{-})/\sqrt{2}$ is the `classical'/`quantum'  component of the fast field $X$ following the Keldysh notation, and $X_{+}(t)$ and $X_{-}(t)$ are their values on the forward and backward time contours. 
Notations for $\phi$ and other symbols are defined in the same way.
Without the external field $f(s)$, one have $Z=1$. However, $Z\neq 1$ if $f(s)$ has different values on the forward and backward time contours so that the path integrals on the two contours don't cancel each other, or in other words, $\hat{U}(s) \neq \hat{I}$.
The functional dependence of $Z$ on $f(s)$ contains the information of observables, correlation functions, etc.

The Keldysh action for a generic periodically driven system (\equa{eqn:action}) is
\begin{align}
	S
	=&S_f[X_c, X_q; \phi_c, \phi_q] + S_s[\phi_c, \phi_q] 
	+  \int dt
    f(t)
	\big(
	P[X_+,\phi_+]-P[X_-,\phi_-]
	\big)
	,\quad
	f(t)=2f  \cos(\omega t) 
	\,.
	\label{eqnSI:action}
\end{align}
\equa{eqnSI:action} is, in general, the action after integrating out any degrees of freedom from the bath.
Further integrating out the fast degree of freedom $X$ in \equa{eqnSI:Z}, one obtains the effective action for $\phi$:
\begin{align}
	S[\phi_c, \phi_q; f] 
	= S_{\text{s}}[\phi_c, \phi_q] 
	+ S_{\text{P}}[\phi_c, \phi_q; f] 
	\,, \quad
	S_{\text{P}} = 
	-\int dt	\phi_q F_{\text{P}}(\phi_c, f)
	+ O(\phi_q^2)
	\,.
	\label{eqnSI:action_phi}
\end{align}
From the construction of the Keldysh action,  the coefficient of the $O(\phi_q)$ term is the derivative of the  potential for $\phi$. Therefore, $F_{\text{P}}(\phi_c, f)$ is the ponderomotive force and $V_{\text{P}}(\phi_c, f)=-\int d\phi_c F_{\text{P}}(\phi_c, f)$ is the ponderomotive potential. In the following, we prove \equa{eqn:lemma1}  for the two cases stated there.
\\

\subsection{Response functions}
In this subsection, we define the linear and nonlinear response functions of the generalized polarization $P$ to the generalized force $f(t)$ in \equa{eqnSI:action}~\cite{Altland.2010, Kamenev.book}. We allow the force (source field) $f(s)$ to have different values $f_+(t),\, f_-(t)$ on the two contours  and define $f_c(t)$ and $f_q(t)$  in the  same way as the fields $X$ and $\phi$, so that the coupling term in \equa{eqnSI:action} generalizes to $f_c P_q+ f_q P_c$ where $	P_{c,q}=(P[X_+,\phi_+] \pm P[X_-,\phi_-])\sqrt{2}$. The physical external field has $f_c(t)$ only, while $f_q(t)$ is introduced for convenience.
The physical polarization is found from the generating functional $Z=e^{-i S[f_c, f_q]}$ as: 
\begin{align}
P_c(t) &= -i\frac{\delta Z[f_c, f_q]}{\delta f_q(t)} \Big|_{f_q=0}
= \sum_{m=1}^{\infty} 
\int dt_1 dt_2 ... dt_m
\frac{1}{m!}  \chi_{\text{r}}^{(m)}(t; t_1, t_2, ..., t_m)
\cdot f_c(t_1) f_c(t_2)... f_c(t_m)
\,.
	\, 
	\label{eqnSI:response1}
\end{align}
Here
\begin{align}
 \chi_{\text{r}}^{(m)}(t; t_1, t_2, ..., t_m)\equiv
-i \frac{\delta Z[f_c, f_q]}{\delta f_q(t) \delta f_c(t_1) \delta f_c(t_2)...\delta f_c(t_m)} \Big|_{f_{c,q}=0}
 =\langle 
 P_c(t) P_q(t_1) P_q(t_2)...P_q(t_3)
 \rangle
\, 
\label{eqnSI:response2}
\end{align}
is defined as the $m$-th order symmetrized retarded response function and $\langle ...\rangle$ means the correlation function under the Keldysh path integral at $f_{c,q}=0$ in \equa{eqnSI:Z}. Therefore, the effective action for $f$  could be written as
\begin{align}
S[f_c, f_q] 
	= -\sum_{m=1}^{\infty} 
	\int dt_1 dt_2 ... dt_m
	\frac{1}{m!}  \chi_{\text{r}}^{(m)}(t; t_1, t_2, ..., t_m)
	\cdot f_q(t) f_c(t_1) f_c(t_2)... f_c(t_m)
	\,.
	\, 
	\label{eqnSI:Sf}
\end{align}
at order $f_q$. If the source field is a single frequency one, meaning $f_{c/q}(t)=2f_{c/q} \cos(\omega t)$, the effective action contains the products of $f$ whose frequencies sum to zero. This occurs for odd $m=2n-1$ only: 
\begin{align}
	S[f_c, f_q] 
	= -\sum_{n=1}^{\infty} 
	\frac{C_{2n}^n}{(2n-1)!}  \chi_{\text{r}}^{(2n-1)}(\omega, -\omega, \omega, ...)
	\cdot f_q f_c^{2n-1}
	,\quad 
	 \chi_{\text{R}}^{(2n-1)}  \equiv  \frac{1}{(n!)^2} \chi_{\text{r}}^{(2n-1)}
	\label{eqnSI:Sf_freq}
\end{align}
where $\chi_{\text{R}}^{(2n-1)}(\omega, -\omega, \omega, ...)$ is  the response function \equa{eqnSI:response2} in the frequency representation,  and the $C_{2n}^n$ factor comes from the different ways of assigning the $n$ positive frequencies ($\omega$) and  $n$ negative frequencie ($-\omega$) to $f_{c/q}$. We have also defined the new response function $\chi_{\text{R}}^{(2n-1)}$ in \equa{eqnSI:Sf_freq} for notational simplicity of the main text.

\subsection{Case 1}
Since there is no dissipation, there is no bath that is integrated out, so that the system is an isolated system. Therefore, the fields on the forward and backward time contours are not coupled~\cite{Kamenev.book,Altland.2010}, meaning the action in \equa{eqnSI:action} could be written as $S=S_+-S_-$ where 
\begin{align}
	S_+ &= \int dt L[X_+, \phi_+,f_+] 
	,\quad
	S_-=\int dt L[X_-, \phi_-,f_-] 
	,\quad \notag\\
	L &=L_f[X,\phi] + L_s[\phi] +  2f  \cos(\omega t) \cdot P[X,\phi]
	\,
	\label{eqnSI:action_case1}
\end{align}
and $f_+=f_-=f$. However, we have given different labels to $f$ on the two contours for later convenience.
To derive the $F_{\text{P}}$ in \equa{eqnSI:action_phi}, one just needs to assume  constant $\phi_+$ and $\phi_-$ on the two contours.
Because we assume that the NESS exists, the path integral on each contour is a  product of infinite periods $T$ which may or may not be $2\pi/\omega$. In the limit of $t_f \rightarrow \infty$, the boundary conditions at $t=0$ and $t=t_f$ that glue the actions on the two contours will not be important, so that the two path integrals over $X$ on the two contours are decoupled from each other. Therefore, after integrating out $X$, one has $S=S_+-S_-$ where
\begin{align}
	S_+ &= \sum_{n=1}^\infty \chi^{(2n-1)}(\phi_+) f_+^{2n}
	,\quad
	S_-= \sum_{n=1}^\infty \chi^{(2n-1)}(\phi_-) f_-^{2n}
	\,.
	\label{eqnSI:action_phi_case1}
\end{align}
For $\phi_+=\phi_-=\phi$ in \equa{eqnSI:action_phi_case1}, one has
\begin{align}
	S&=S_+-S_-
	=
	\sum_{n=1}^\infty \chi^{(2n-1)}(\phi) (f_+^{2n} -f_-^{2n})
	=\sum_{n=1}^\infty \chi^{(2n-1)}(\phi) 
	\left( 2n f_q f_c^{(2n-1)} + O(f_q^3)
	\right)
	\,.
	\label{eqnSI:case1_1}
\end{align}
Since the coefficient of the $O(f_q)$ term should be the retarded response functions, comparing with \equa{eqnSI:Sf_freq} leads to $\chi^{(2n-1)}(\phi)= -\chi_R^{(2n-1)}(\phi)$. 
Now, setting $f_+=f_-=f$ in \equa{eqnSI:action_phi_case1}, one has
\begin{align}
	S&=S_+-S_-=
	\sum_{n=1}^\infty 
	\left(\phi_q \partial_{\phi_c} \chi^{(2n-1)}(\phi_c) + O(\phi_q^3) \right)
	f^{2n} 
	\,.
	\label{eqnSI:case1_2}
\end{align}
The above two equations combined with \equa{eqnSI:action_phi} lead to the conclusion that the ponderomotive force is $F_{\text{P}}(\phi_c)= \sum_n \partial_{\phi_c} \chi_R^{(2n-1)}(\phi_c) f^{2n} $ and the ponderomotive potential is $V_{\text{P}}(\phi_c)= -\sum_n  \chi_R^{(2n-1)}(\phi_c)  f^{2n} $.
\\

\subsection{Case 2}
Case~2 can incorporate any bath (integrated out) and  dissipation in \equa{eqnSI:action}. However, it requires $S_f =S_f[X_c, X_q]$ such that the `fast' action for $X$ is not affected by $\phi$. It also requires that the generalized polarization  $P$ could be separated as a product $P_1(X) P_2(\phi)$ in \equa{eqn:action}. Up to the linear order in $\phi_q$, the coupling Lagrangian in \equa{eqnSI:action} could be written as
\begin{align}
	L_c &=
	2f  \cos(\omega t) \cdot 
	\left[
	P_q(X_c,X_q,\phi_c)  + \phi_q	\partial_{\phi_c} P_c(X_c,X_q,\phi_c)
	\right]
	\,, \notag\\
	P_c &=
	P[X_c+X_q,\phi_c]+P[X_c-X_q,\phi_c]
	\,,\quad 
	P_q =
	P[X_c+X_q,\phi_c]-P[X_c-X_q,\phi_c]
	\,,
	\label{eqnSI:P}
\end{align}
where $P_c$ and $P_q$ are the `classical' and `quantum' components of the generalized polarization. From \equa{eqnSI:action_phi}, after integrating out $X$, the ponderomotive force for $\phi_c$ is therefore
\begin{align}
	F_{\text{P}}(\phi_c) &= 
	\sum_{n=1}^\infty  		\frac{C_{2n}^n}{(2n-1)!}   C^{(2n)}(\phi_c) f^{2n}
	\,,\quad 
	C^{(2)}(\phi_c) =
	\langle P_q \partial_{\phi_c} P_c\rangle|_{(\omega, -\omega)}
	\,,\quad 
	C^{(4)}(\phi_c) =
	\langle P_q  P_q P_q  \partial_{\phi_c} P_c\rangle|_{(\omega, -\omega,\omega,-\omega)}
	\,,... \,,
	\label{eqnSI:P_force_case2}
\end{align}
where $\langle \rangle$ means the correlation function: functional average over $X_c(t), X_q(t)$  using the Keldysh path integral in \equa{eqnSI:Z} at $\phi_q=0$ and fixed $\phi_c$.
Since $P=P_1(X)P_2(\phi)$ is separable, one has $P_c=P_{1c} P_2(\phi_c)$ and $P_q=P_{1q} P_2(\phi_c)$ where $P_{1c/q}=(P_1[X_+] \pm P_1[X_{-}])/\sqrt{2}$. Therefore, for constant $\phi$, the coefficients can be written as 
\begin{align}
	C^{(2)}(\phi_c) &=
	\frac{1}{2}\partial_{\phi_c}	\langle P_q  P_c\rangle|_{(\omega, -\omega)}
	=\frac{1}{2} \partial_{\phi_c} \chi_r(\omega, -\omega)
	\,,\quad
	\notag\\
	C^{(4)}(\phi_c) &=
	\frac{1}{4}	\partial_{\phi_c}  
	\langle P_q  P_q P_q  P_c\rangle|_{(\omega, -\omega,\omega,-\omega)}
	= \frac{1}{4}		\partial_{\phi_c}  
	\chi_r^{(3)}|_{(\omega, -\omega,\omega,-\omega)}
	\,,... \,,
	\label{eqnSI:C_case2}
\end{align}
By comparison with \equa{eqnSI:Sf_freq}, one concludes that $F_{\text{P}}(\phi_c)= \sum_n \partial_{\phi_c} \chi_R^{(2n-1)}(\phi_c) f^{2n}$. Making use of the symmetric  properties of the real parts of response functions, one obtains \equa{eqn:lemma1}.
\\

\section{Light induced exciton condensate}
\label{apdx:condensate}
For  reader's convenience, we reproduce the Lagrangian (\equa{eqn:exciton_Lagrangian_2}) of the excitons here:
\begin{align}
	L =   L_s[\Phi] 
	+ \rho L_\xi
	\,,\quad
	L_s[\Phi] =		\Phi^\ast
	\left( -i\partial_t+ \omega_{\text{ex}}  
	\right)  \Phi
	+ g \rho^2 
	\,,\quad
	L_\xi =
	\eta (\dot{\varphi} + \omega_0)
	+ \lambda E(t) \sqrt{\eta(1-\eta)} 2 \cos \varphi
	\,.
	\label{eqnSI:exciton_Lagrangian_2}
\end{align}
Writing the internel degree of freedom $\phi, \, \eta$ as a pseudo-vector on the unit sphere:
\begin{align}
	\vec{n}=\xi \vec{\sigma} \xi^{\dagger}=\left( 2\sqrt{\eta(1-\eta)} \cos \varphi,\, - 2\sqrt{\eta(1-\eta)} \sin \varphi,\, 1-2\eta \right)
\end{align}
$L_\xi$ could also be written as 
\begin{align}
	L_\xi=-\vec{A}(\vec{n}) \cdot \dot{\vec{n}} 
	-\vec{B}(t) \cdot  \vec{n} + B_z
	\,,\quad
	(B_x,\, B_y,\, B_z)= \left(\lambda E(t),\, 0,\, \frac{\omega_0}{2} \right)
\end{align}
where $\vec{B}$ is the effective `Magnetic field' and $\vec{A}(\vec{n})$ is the Berry connection on the unit sphere for the spin coherent state path integral. The corresponding Berry curvature  is $\vec{\Omega}(\vec{n})=d\vec{A}(\vec{n})=\vec{n}/2$.
After adding the dissipative terms, one may write down the Keldysh version of \equa{eqnSI:exciton_Lagrangian_2}:
\begin{align}
	L &=  L_s[\Phi_c, \Phi_q] + L_f[\Phi_c, \Phi_q, \vec{n}_c, \vec{n}_q]
	,\quad
	\notag\\
	L_f &=
	\Phi_c^\ast  \Phi_c \vec{n}_q \cdot
	\bigg[ -\dot{\vec{n}}_c \times \vec{\Omega}(\vec{n}_c)
	-\vec{B}(t)
	+\gamma  \dot{\vec{n}}_c 
	\bigg]
	+ \left[ \Phi_q^\ast  \Phi_c 
	\left[ -\dot{\vec{n}}_c \vec{A}(\vec{n}_c)
	-\vec{B}(t)\vec{n}_c + B_z
	\right] +c.c.
	\right]
	\,.
	\label{eqnSI:exciton_S}
\end{align}
The classical saddle point $\partial_{n_q} L =0$ for the spin (together with the constraint that $|\vec{n}|=1$)  is just the Landau–Lifshitz–Gilbert equation for a classical magnet:
\begin{align}
	\dot{\vec{n}}_c = 2\vec{n}_c \times \left(
	-\vec{B}(t)
	+\gamma  \dot{\vec{n}}_c 
	\right)
	\,.
	\label{eqnSI:LLG}
\end{align}
It is has the interpretation of a massless particle moving on the unit sphere under the magnetic field 
$\vec{\Omega}(\vec{n}_c)$ and in the potential $-\vec{B} \cdot \vec{n}_c$,  and experiencing the friction  $-\gamma \dot{\vec{n}}_c$.

\subsection{Stable orbit of the driven classical spin}
\begin{figure}
	\includegraphics[width=  \linewidth]{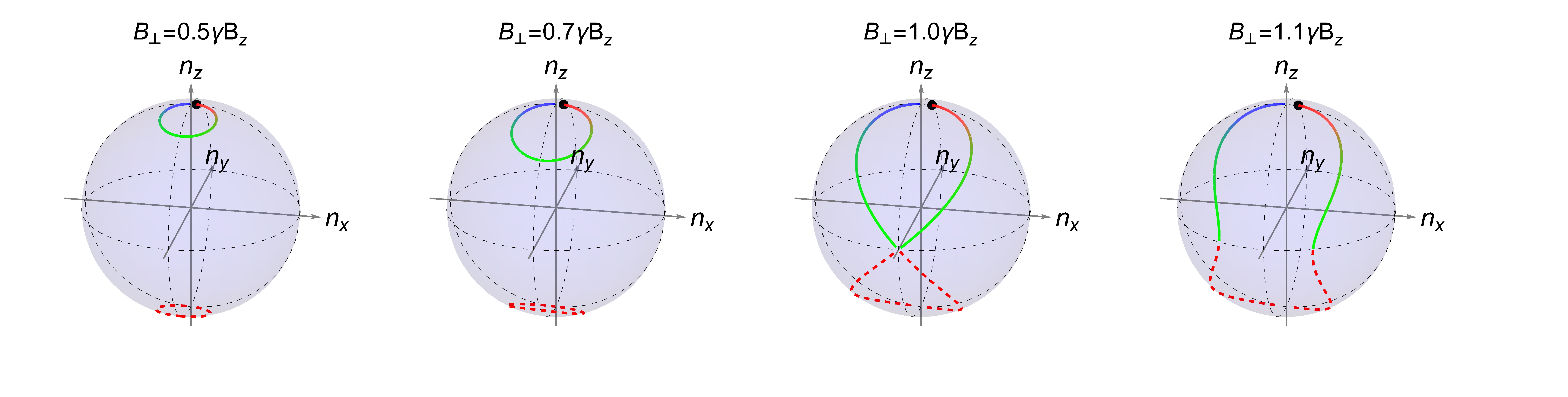} 
	\caption{Solid curves are the stable stationary $\vec{n}_c=\left( \sin \theta_0 \cos\varphi_0,\,   \sin \theta_0  \sin\varphi_0 ,\, \cos \theta_0
		\right) $ of the driven dissipative classical spin in the rotating frame as a function of the driving frequency $\omega$. On these curves, red means $\omega=0$, green means $\omega=\omega_0$, blue means $\omega = \infty$ and other values of frequency interpolates among these colors. The black dots are the stable points at $\omega=0$. The red dashed curves are the corresponding unstable stationary points. The four panels are for different values of the driving field $B_\perp=\lambda E_0/2$. The dimensionless damping rate is $\gamma=0.1$. Beyond the critical driving field, $B_\perp=\gamma B_z>\gamma \omega_0/2$, the curve becomes disconnected, meaning that there is a jump of $\vec{n}_c$ as the driving frequency $\omega$ crosses $\omega_0$.}
	\label{figSI:driven_spin}
\end{figure}

To compute the ponderomotive force for $\Phi$, one needs to solve for the stable classical orbit of $\vec{n}_c$ from  \equa{eqnSI:LLG}, for which a general analytical solution is unknown. Fortunately, since we aim at  the resonance regime $|\omega-\omega_0| \lesssim \lambda E_0 $ to obtain the higher order corrections to the $O(E_0^2)$ result, it is reasonable to make the rotating wave approximation: 
\begin{align}
	(B_x,\, B_y,\, B_z)=
	\left(\lambda E_0 \cos (\omega t),\, 0,\, \frac{\omega_0}{2} \right)
	\rightarrow 
	\left( B_\perp \cos (\omega t),\, B_\perp \sin (\omega t),\, B_z
	\right) 
	\,,\quad 
	B_\perp =  \frac{\lambda E_0}{2}
	\,,\quad 
	B_z = \frac{\omega_0}{2} 
	\,.
\end{align}
Now the drive is an effective magnetic field $B_\perp$ rotating on the $x-y$ plane with angular  frequency $\omega$. The stable orbit should be the unit vector $\vec{n}_c(t)
=\left( \sin \theta_0 \cos(\omega t +\varphi_0),\,   \sin \theta_0  \sin(\omega t +\varphi_0),\, \cos \theta_0
\right) $  rotating at the same angular  frequency. In the rotating frame,  it is a stable point $\vec{n}_c=\left( n_{x},\, n_{y},\,n_{z}\right) = \left( \sin \theta_0 \cos\varphi_0,\,   \sin \theta_0  \sin\varphi_0,\, \cos \theta_0
\right) $. Plugging it into \equa{eqnSI:LLG} yields the equation for the angles:
\begin{align}
	-\delta \sin \theta_0 -B_\perp \cos \theta_0 \cos \varphi_0 =0
	\,,\quad
	\frac{1}{2}\gamma \omega \sin \theta_0  +  B_\perp \sin \varphi_0   = 0	
	\,.
\end{align}
where $\delta= (\omega- \omega_0)/2$ is the detuning. The solution is 
\begin{align}
	n_{z}^2=1- \frac{1}{2} \left[
	1+ \frac{1}{\gamma^2 \omega^2/4} \left( \delta^2 + B_\perp^2 \right)
	\pm 
	\sqrt{
		\left(1+ \frac{1}{\gamma^2 \omega^2/4} \left( \delta^2 + B_\perp^2 \right)
		\right)^2
		-4  \frac{B_\perp^2}{\gamma^2 \omega^2/4} 
	}
	\right]
	\,,\quad
	\sin \varphi_0 = -\frac{\gamma \omega/2}{B_\perp} \sin\theta_0
	\,.
	\label{eqnSI:spin_solution}
\end{align}
Note that for each  field $B_\perp$ and driving frequency $\omega$, there is a pair of solutions to \equa{eqnSI:spin_solution}, i.e., $(\theta_0, \varphi_0)$ and $(\pi-\theta_0, \pi-\varphi_0)$. These two points as functions of the driving frequency $\omega$ correspond to the solid and dashed curves in \fig{figSI:driven_spin}. However, only the point on the north hemisphere (solid curves) is the stable one, while that on the south hemisphere (dashed curves) is unstable which would relax to the former location due to fluctuations. From the  perspective  of continuity, as the driving field increases gradually from zero to beyond the critical one, the solid curve continuously expands from the circle in the left panels to the disconnected curves in the right panel. 

\subsection{The ponderomotive potential}
We now compute the ponderomotive force $F_{\text{P}}(\Phi_c) = \lim_{T\rightarrow \infty} \langle \frac{1}{T} \int_0^T dt \partial_{\Phi_q^\ast}  L_f \rangle$ for $\Phi$ at the mean field level. This means we approximate the path integral average $\langle \rangle$ by the classical saddle point from the previous section: 
\begin{align}
	F_{\text{P}}(\Phi_c) = -\langle \partial_{\Phi_q^\ast} L_f \rangle_t
	= -\Phi_c  \langle 
	\left[ -\dot{\vec{n}}_c \vec{A}(\vec{n}_c)
	-\vec{B}(t)\vec{n}_c
	+B_z
	\right]  
	\rangle_t
	\,.
	\label{eqnSI:exciton_condensate_P_force}
\end{align}
Plugging the stable orbit of \equa{eqnSI:spin_solution} into \equa{eqnSI:exciton_condensate_P_force} yields the ponderomotive force, and the ponderomotive potential
\begin{align}
	V_{\text{P}}(\Phi)=|\Phi|^2  \langle 
	\left[ -\dot{\vec{n}}_c \vec{A}(\vec{n}_c)
	-\vec{B}(t)\vec{n}_c+B_z
	\right]  
	\rangle_t
	=|\Phi|^2  
	\left[ 
	\left(B_z-\omega/2\right)(1-n_{z})
	-B_\perp n_{x}
	\right]  
\end{align} 
used for \fig{fig:exciton_phase_diagram}. At the resonant  $\omega \rightarrow \omega_0$, one has 
\begin{align}
	(n_x,\, n_z ,\, V_{\text{P}})
	=	
	\left\{
	\begin{array}{lc}
		\left(
		0
		,\quad
		\sqrt{1-\frac{B_\perp^2}{\gamma^2 B_z^2}} 
		,\quad
		0
		\right)
		\,,\quad
		& B_\perp \leq  \gamma B_z
		\\
		\left( 
		\sqrt{1-\frac{\gamma^2 B_z^2}{B_\perp^2}} 
		,\quad 1
		,\quad
		-|\Phi|^2  B_\perp
		\sqrt{1-\frac{\gamma^2  B_z^2 }{B_\perp^2}} 
		\right) \,,\quad
		& B_\perp >  \gamma B_z \,\, \&  \,\,  \omega=\omega_{0-}
		\\
		\left( 
		-\sqrt{1-\frac{\gamma^2 B_z^2}{B_\perp^2}} 
		,\quad 1
		,\quad
		-|\Phi|^2  B_\perp
		\sqrt{1-\frac{\gamma^2  B_z^2 }{B_\perp^2}} 
		\right) \,,\quad
		& B_\perp >  \gamma B_z \,\, \&  \,\,  \omega=\omega_{0+}
	\end{array}
	\right.
	\label{eqnSI:spin_resonance}
\end{align}
where $B_\perp=\lambda E_0/2=c_p a_0 eE_0$ is the energy scale of the field strength. For a dielectric screening of $\epsilon=7.0$, one obtains the Bohr radius $a_0=0.74 \unit{nm}$, the $s \rightarrow p$ transition energy $\omega_0=104 \unit{meV}$  and the shape factor $c_p=0.53$. These parameters give $B_\perp =2 \unit{meV}$ for an electric field of $E_0=10^5 \unit{V/cm}$. \fig{fig:exciton_phase_diagram}   is based on these parameters.

\section{Light enhanced electron-electron attraction}
\label{apdx:SC}
\subsection{The infrared phonon model}
\label{apdx:IR_phonon}

In this section, we consider the infrared phonon coupled to a general inversion-even electronic degree of freedom $\phi$:
\begin{align}
	L =&  \frac{1}{2} 
	\left[- \dot{X}^2 + (\omega_0^2 + \phi) X^2  
	\right]
	+ E(t) X 
	\,.
	\label{eqnSI:electron_IR_phonon}
\end{align}
Note that $\phi$ is in general a functional of electronic degrees of freedom which may also contain time derivatives, and reduces to $\phi=g_e \rho$ in the simple case discussed in section~\ref{sec:sc}. After adding the damping rate $\gamma$ of the phonon, the Keldysh version~\cite{Kamenev.book,Altland.2010} of the Lagrangian is written as 
\begin{align}
	L = \frac{1}{2}
	\begin{pmatrix}
		X_c &
		X_q 
	\end{pmatrix}
	\begin{pmatrix}
		\phi_q  & \partial_t^2 + \gamma \partial_t + \omega_0^2 + \phi_c \\
		\partial_t^2 - \gamma \partial_t + \omega_0^2 +  \phi_c    &  
		- 2\gamma \partial_t \coth \frac{i\partial_t}{2T} +\phi_q 
	\end{pmatrix}
	\begin{pmatrix}
		X_c \\
		X_q 
	\end{pmatrix}
	+ 
	E(t) X_q
	\label{eqnSI:IR_phonon_action}
\end{align}
to linear order in $\phi_q$, where $T$ is the temperature of the bath. Integrating out $X$ and looking for the $O(\phi_q)$ term according to \equa{eqnSI:Z}, one obtains the exact ponderomotive force for $\phi$:
\begin{align}
	F_{\text{P}} = - \frac{1}{2} \langle 
	X^2
	\rangle 
	= - \frac{1}{2} \left(X_0^2 + X^2_{\text{driven}} \right)
	=
	F_0+ 
	\chi_R(\phi_c, \omega)
	\chi_A(\phi_c, \omega)
	E_0^2
	\,,\quad
	\chi_R(\phi_c, \omega) = \frac{1}{-\omega^2 - i\gamma \omega + \omega_0^2 + \phi_c}
	\,,\quad
	\chi_A=\chi_R^\ast
	\,
	\label{eqnSI:IR_phonon_fp}
\end{align}
where $\chi_R$ and $\chi_A$ are the retarded and advanced response functions of the IR phonon. 

Note that $F_0$ is not induced by the drive, but is the  contribution to $F_{\text{P}}$ arising from the equilibrium quantum/thermal fluctuations in $X$:
\begin{align}
	F_0= - \frac{1}{2} X_0^2
	= - \frac{1}{2} \frac{E_T(\phi, T)/a^3}{\omega_0^2 + \phi}
	\,,\quad
	E_T(\phi, T) =  \hbar \sqrt{\omega_0^2 + \phi} 
	\left[
	n_b\left( \hbar \sqrt{\omega_0^2 + \phi},\, T \right) + \frac{1}{2}
	\right]
	\,
	\label{eqnSI:F0}
\end{align}
where $E_T$ is the energy of the local phonon, $a$ is the lattice constant, $n_b$ is the boson occupation number, and we have restored $\hbar$.
At low temperatures $T \ll \omega_0$, the equilibrium contribution to the ponderomotive force in \equa{eqnSI:F0} is just from the quantum fluctuations (zero point motion) of $X$.
The resulting ponderomotive force/potential is
\begin{align}
	F_0
	= - \frac{\hbar}{4a^3}
	\frac{1}{ \sqrt{\omega_0^2 + \phi} }	
	\,,\quad
	V_0
	= \frac{\hbar}{2a^3}
	\sqrt{\omega_0^2 + \phi}
	\,.
	\label{eqnSI:V0}
\end{align}

The drive induced ponderomotive potential is
\begin{align}
	V_{\text{P}}(\phi) &= 
	\frac{E_0^2}{\gamma \omega}
	\arctan
	\frac{\gamma \omega}{\omega^2-(\omega_0^2+ \phi)}  
	\,.
	\label{eqnSI:vp_ep_damping}
\end{align}
Setting $\phi= g_e \rho$ in \equa{eqnSI:vp_ep_damping} renders \equa{eqn:vp_ep_damping}.

\subsubsection{Estimations for K$_3$C$_{60}$}
\label{apdx:k3c60}
In this section, we make rough estimations of possible light induced superconductivity in the Fulleride K$_3$C$_{60}$. The dimensionless strength of the local density-density interaction is 
\begin{align}
	\nu g_{\text{P}} &=	\nu \frac{ \omega^2-\omega_0^2}{\left[(\omega^2-\omega_0^2)^2+\gamma^2 \omega^2 \right]^2}
	g_e^2 	E_0^2 
	= \nu  
	X_{\text{max}}^2
	\frac{\omega_0^2 \gamma^2- \gamma^4/4}{(\omega^2-\omega_0^2)^2+\gamma^2 \omega^2 }
	\frac{ \omega^2-\omega_0^2}{(\omega^2-\omega_0^2)^2+\gamma^2 \omega^2 }
	g_e^2
	\,
	\label{eqnSI:gp}
\end{align}
where $\nu$ is the density states of the electrons at the Fermi energy, and $X_{\text{max}}^2= \frac{E_0^2}{\omega_0^2 \gamma^2- \gamma^4/4}$ is the maximum mean square displacement of the phonon that occurs at the frequency $\omega^2=\omega_0^2-\gamma^2/2$. 

We first estimate $X_{\text{max}}$.
Although the pump field $E_0$ could be very strong, to ensure that the lattice is not destroyed, the maximum possible displacement $x$ of the $T_{1u}$ mode is about $\kappa \sim 0.1$ times the lattice constant $a$. The corresponding Kinetic energy of the local Harmonic oscillator is $K \approx  \frac{m_0}{4} \dot{x}^2 \sim \frac{1}{4} m_0 \omega_0^2 (\kappa a)^2$ where $m_0$ is the mass of a $K_3 C_{60}$ molecule. Since the kinetic energy is also equal to $\frac{1}{2} \omega_0^2 X_{\text{max}}^2 a^3$ according to the continuous field theory in \equa{eqnSI:electron_IR_phonon}, we arrive at $X_{\text{max}}^2 \sim m_0 \kappa^2 /a$.

Now we estimate $g_e$ from Ref.~\cite{Giannozzi.1996} following Ref.~\cite{Kennes2017}. Experimentally~\cite{Giannozzi.1996} , the $T_{1u} (1)$ phonon (the lowest $T_{1u}$ mode) 
of C$_{60}$ with the frequency $\omega_2 = 16 \unit{THz}$ is shifted down by  about $2 \unit{THz}$  to $\omega_1 = 466 \unit{cm}^{-1} = 14 \unit{THz}$  after being doped to 
K$_6$C$_{60}$. Assuming the downshift is only through the coupling $g_e$  to the  doped electrons with density $\rho_0$, the coupling constant is estimated as $g_e \sim (\omega_2^2-\omega_1^2)/\rho_0$.

Now \equa{eqnSI:gp} could be written as 
\begin{align}
	\nu g_{\text{P}} &\sim \nu  
	\kappa^2 \frac{m_0}{a}
	\frac{\omega_0^2 \gamma^2- \gamma^4/4}{(\omega^2-\omega_0^2)^2+\gamma^2 \omega^2 }
	\frac{ \omega^2-\omega_0^2}{(\omega^2-\omega_0^2)^2+\gamma^2 \omega^2 }
	\frac{(\omega_2^2-\omega_1^2)^2}{\rho_0^2}
	\notag\\
	&\sim
	\kappa^2 
	\frac{\omega_0^2 \gamma^2- \gamma^4/4}{(\omega^2-\omega_0^2)^2+\gamma^2 \omega^2 }
	\frac{ (\omega^2-\omega_0^2) (\omega_2^2-\omega_1^2)}{(\omega^2-\omega_0^2)^2+\gamma^2 \omega^2 }
	\frac{m_0 a^2(\omega_2^2-\omega_1^2)}{ W }
	\,.
	\label{eqnSI:gp2}
\end{align}
In the second equality we have made use of $\nu \sim \rho_0/W$ where $W$ is the band width, and $\rho_0 \sim 1/a^3$. 

The equilibrium contribution  $g_0$ to the local interaction at low temperatures can be found by expanding \equa{eqnSI:V0} to $\rho^2$, which gives
\begin{align}
	\nu g_0
	= -\frac{\hbar}{16a^3} \frac{\nu g_e^2}{\omega_0^3}
	= -\frac{\hbar}{16a^3} \frac{\rho_0/W}{\omega_0^3}   \frac{(\omega_2^2-\omega_1^2)^2}{\rho_0^2}
	=-\frac{\hbar}{16} \frac{(\omega_2^2-\omega_1^2)^2}{W \omega_0^3}  
	\ll 1
	\,.
	\label{eqnSI:V0}
\end{align}

The black curve in \fig{fig:superconductivity_phase_diagram} is plotted as $\nu(g_P+g_0)$ from Eqs.~\eqref{eqnSI:gp2} and \eqref{eqnSI:V0} with $\kappa = 0.015$, $\omega_2=11 \unit{THz}$, $\omega_1 = 10\unit{THz}$, $\omega_0 = 10\unit{THz}$, $\gamma= 2\unit{THz}$, $W = 1 \unit{eV}$, $a= 1 \unit{nm}$ and  $m_0\approx 1.53 \times 10^6 m_e$. The maximum attraction occurs at about $\omega = \omega_0^2 - \gamma \omega_0$ which yields 
$|\nu g_P| \sim  \kappa^2 \frac{\omega_2^2-\omega_1^2}{\gamma \omega_0 }
\frac{m_0 a^2(\omega_2^2-\omega_1^2)}{ W }
\sim 1
$. 

\subsubsection{Light induced heating}
\label{apdx:heating}
We now estimate the heating effect in a recent pump probe experiment~\cite{Rowe.2023_k3c60} on K$_3$C$_{60}$. To show a typical estimate, we pick the experimental parameters at the base temperature of $T_{\text{0}}=100 \unit{K}$ (Fig.~3 of Ref.~\cite{Rowe.2023_k3c60}). The pump pulse has the central frequency $\omega=10 \unit{THz}$, the duration $1 \unit{ps}$ and the  fluence  $f_{\text{E}}=0.5 \unit{mJ/cm^{2}}$, corresponding to a peak electric field of $E_0=1.23 \times 10^4 \unit{V/cm}$. Assuming the pumped layer is heated up from the base temperature $T_{\text{0}}$ to a well defined temperature $T_{\text{0}}+\delta T$, the increase in temperature is estimated as 
\begin{align}
	\label{eqn:dletaT}
	\delta T=\frac{f_{\text{E}}}{C_v d} \approx 1.5 \unit{K}
\end{align}
where $C_v$ is the heat capacity, $d=2\pi/\mathrm{Im}[\sqrt{\epsilon}\omega/c]$ is the penetration depth of light at the frequency $\omega=10 \unit{THz}$, $\epsilon$ is the dielectric at this frequency, and $c$ is the speed of light. In Ref.~\cite{Allen1999}, the specific heat of K$_3$C$_{60}$ is measured to be $\approx 0.2 \unit{J/(gK)}$ at 100 K, equivalent to a heat capacity of $C_v\approx 0.39 \unit{J/(cm^3 K)}$ considering that it has a molar mass of $837 \unit{g/mol}$ and a lattice constant of $14.175 \unit{\AA}$~\cite{Rowe.2023_k3c60}. The dielectric is $\epsilon=1+\frac{4\pi i}{\omega} \sigma(\omega)=-4.4 + 36.0 i$ from the optical conductivity $\sigma(\omega)=(200+ 30 i ) \unit{\Omega^{-1} cm^{-1}}$ measured at $\omega=10 \unit{THz}$ (see Fig.~3b in Ref.~\cite{Rowe.2023_k3c60}), giving  a  penetration depth of $d=6.6 \unit{\mu m}$. 
If the pulse duration is increased to $4 \unit{ps}$ (containing $40$ cycles of field oscillation) such that it is convenient to perform a Floquet measurement, the system is heated up by just $6 \unit{K}$. 

A similarly estimate for Ta$_2$NiSe$_5$ renders $\delta T \approx 16 \unit{K}$ given the same fluence $f_{\text{E}}=0.5 \unit{mJ/cm^{2}}$, the pumping frequency $\omega=0.2 \unit{eV}$,  the base temperature $T_{\text{0}}=150 \unit{K}$, and the heat capacity and optical properties measured in Ref.~\cite{Lu2017}.

We note that the above estimates are for time regimes after electron-phonon heat transfer (which finishes typically within picoseconds) so that the large heat capacity of the lattice plays a role. Before electron-phonon heat transfer, the injected energy is still in the electronic system, and one may expect the electrons to be much hotter than the estimates. However, not all the energy of the pump is converted into heat. In the  clean systems studied in this paper, the energy of the pump is actually converted into collective degrees of freedom, a nonthermal effect that provides the ponderomotive potential.

\subsection{The Raman phonon model}
\label{apdx:Raman_phonon}

Another example of light induced e-e attraction is from Raman phonons~\cite{Babadi.2017}:
\begin{align}
	L =&  \frac{1}{2} 
	\left[- \dot{X}^2 + \omega_0^2 (1+ \lambda E(t)^2 ) X^2  
	\right]
	+ \rho X 
	\,
	\label{eqnSI:electron_raman_phonon}
\end{align}
where $E(t)^2= (E_0 \cos \omega t)^2$ couples parametrically to  the Raman phonon $X$ either directly or  via a IR phonon.  
\equa{eqnSI:electron_raman_phonon} could be transformed as
\begin{align}
	L =&  
	- \frac{1}{2}  \dot{X}^2 + 
	\frac{1}{2} \omega_0^2 \left[1+ \lambda  \frac{1}{2} E_0^2 (1 + \cos 2\omega t) 
	\right] X^2  
	+ \rho X 
	\notag\\
	=& 
	-  \frac{1}{2}  \dot{X}^{\prime 2} +  \frac{1}{2}  \omega_0^{\prime 2} X^{\prime 2}
	- \frac{1}{2  \omega_0^{\prime 2}} \rho^2 
	+ \frac{1}{2}  \omega_0^{ 2} \lambda  \frac{1}{2} E_0^2   \cos 2\omega t
	\left(X^{\prime}-\frac{\rho}{\omega_0^{\prime 2}} \right)^2
	\notag\\
	=& 
	-  \frac{1}{2}  \dot{X}^{\prime 2} 
	+  \frac{1}{2} \left( \omega_0^{\prime 2} + \frac{1}{4}  \omega_0^{ 2} \lambda  E_0^2   \cos 2\omega t\right)
	X^{\prime 2}
	- 
	\frac{\lambda  E_0^2/2}{1+ \lambda  E_0^2/2} 
	(\cos 2\omega t)
	\rho X^{\prime}
	- \frac{1}{2  \omega_0^{\prime 2}} \rho^2 
	\,
	\label{eqnSI:electron_raman_phonon2}
\end{align}
where we defined the shifted phonon frequency $\omega_0^{\prime 2}= \omega_0^2 (1+ \lambda  E_0^2/2) $ and shifted phonon variable $X^{\prime}=X+\frac{\rho}{\omega_0^{\prime 2}}$.  
The electron density $\rho$ is the slow variable and the  phonon $X$ is the fast one.
At order $E_0^4$, the $\frac{1}{4}  \omega_0^{ 2} \lambda  E_0^2   \cos 2\omega t$ term may contribute to the ponderomotive potential only by the parametric response proportional to  quantum/thermal fluctuations, which is small. 
Therefore, we neglect this term and \equa{eqnSI:electron_raman_phonon2} becomes
\begin{align}
	L 	=& 
	\frac{1}{2}  \left( -\dot{X}^{\prime 2} 
	+   \omega_0^{\prime 2} 
	X^{\prime 2} \right)
	- 
	\frac{\lambda  E_0^2/2}{1+ \lambda  E_0^2/2} 
	(\cos 2\omega t)
	\rho X^{\prime}
	- \frac{1}{2  \omega_0^{\prime 2}} \rho^2 
	\,
	\label{eqnSI:electron_raman_phonon3}
\end{align}
where the second term is now the only dynamical driving term.
Even with dissipation, \equa{eqnSI:electron_raman_phonon3} fits in case~2 in \equa{eqn:lemma1}, and the resulting ponderomotive potential is therefore simply obtained from the retarded response function of a Harmonic oscillator as:
\begin{align}
	V_{\text{P}} (\rho)	&=
	\left[
	\mathrm{Re}\left[\frac{1}{-4\omega^2+\omega_0^2- i 2\gamma \omega} 
	\right]
	\left(	\frac{\lambda  E_0^2/2}{1+ \lambda  E_0^2/2} 
	\right)^2
	- \frac{1}{2  \omega_0^2 (1+ \lambda  E_0^2/2) }
	\right]
	\rho^2
	\notag\\
	&=
	\frac{1}{2  \omega_0^2} 
	\left[
	- 1+ \frac{1}{2} \lambda  E_0^2
	+
	\left(
	- \frac{1}{4} 
	+
	\mathrm{Re}\left[\frac{\omega_0^2/2}{-4\omega^2+\omega_0^2- i2 \gamma \omega} 
	\right]
	\right)
	\lambda^2  E_0^4
	+ O(E_0^6)
	\right]
	\rho^2
	\,.
	\label{eqnSI:vp_raman_phonon_vp}
\end{align}
\equa{eqnSI:vp_raman_phonon_vp} is the static component of the effective interaction derived  in Ref.~\cite{Babadi.2017}.
Therefore, there is a positive contribution at order $E_0^2$, and a negative contribution at order $E_0^4$ if $\omega$ is  red tuned  relative to $\omega_0$. Since the negative $ \lambda^2  E_0^4$ term  is resonantly enhanced as $\omega$ gets close to $\omega_0$ from the red tuned side, it is possible for it to exceed the positive $ \lambda E_0^2$ contribution, and enhance the attraction interaction.

\section{Light engineered free energy landscape}
\label{apdx:CSE}
\begin{figure}
	\includegraphics[width=  \linewidth]{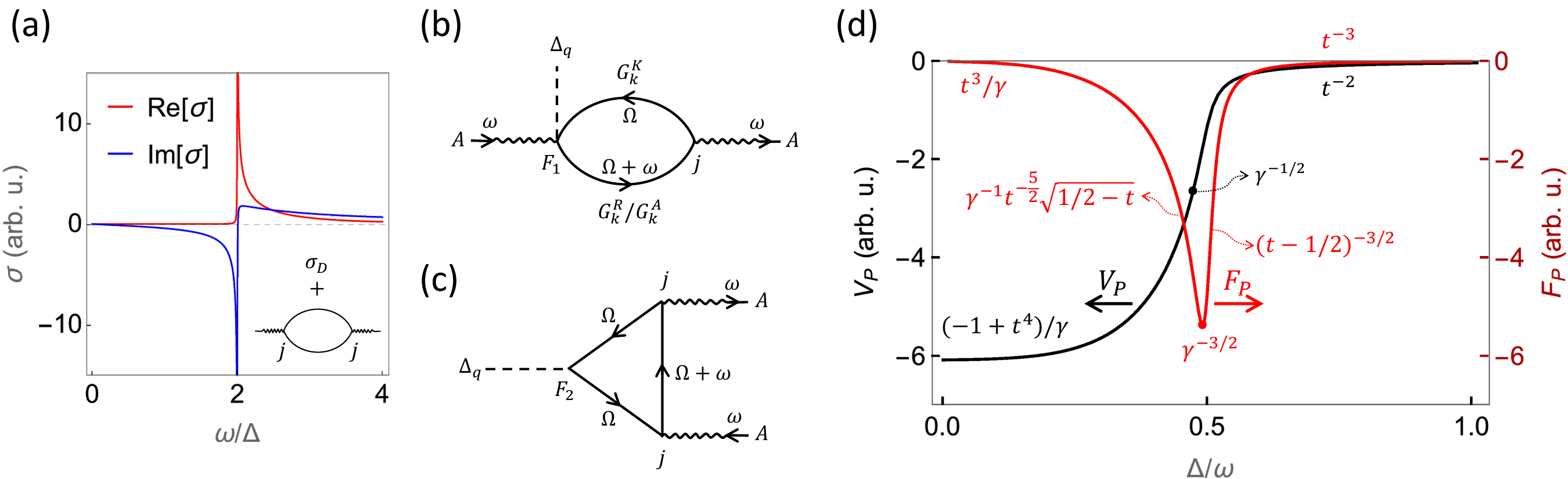} 
	\caption{(a) The  optical conductivity of an electronic system with charge/spin/excitonic order in the BCS weak coupling case. Inset is the contributions to the optical conductivity: the diamagnetic part $\sigma_D= ine^2/(m\omega)$ plus the current-current correlation. (b)/(c) The Feynman diagram  representing the $F_1$/$F_2$ part of the ponderomotive  force. 
		(d) Red curve is the ponderomotive force $F_{\text{P}}(\Delta)$ as a function of $\Delta/\omega$ for a quasi particle damping rate of $\gamma=0.03$ in units of $\omega$. Black curve is the corresponding ponderomotive potential. The inset expressions are their asymptotic behaviors in each region, with $t=\Delta/\omega$, $\gamma$ in units of $\omega$, and the colors matching the curves they refer to. }
	\label{figSI:BCS}
\end{figure}

The Lagrangian for an electronic system exhibiting charge, spin or excitonic order and driven by light $A(t) = A_0(t) \cos \omega t$ is
\begin{align}
	L  &= 
	\int dr	\psi^\dagger 
	\begin{pmatrix}
		-i\partial_t  + \xi(p+A) & \Delta \\
		\Delta^\ast   &  -i\partial_t - \xi(p+A)
	\end{pmatrix}
	\psi
	+ 
	\frac{1}{g} \int dr |\Delta|^2 
	\notag \\
	&= 
	\sum_k
	\begin{pmatrix}
		\psi_{ck}^\dagger 
		&
		\psi_{vk}^\dagger 
	\end{pmatrix}
	\left[
	-i\partial_t + E_k \sigma_3
	+
	\vec{A} \cdot 
	\vec{j}_k
	\right]
	\begin{pmatrix}
		\psi_{ck}
		\\
		\psi_{vk}
	\end{pmatrix}
	+
	\int dr \left( \frac{1}{g} |\Delta|^2  + \frac{n}{m} A^2 \right) 
	\label{eqnSI:BCS_HS_action}
\end{align}
where $\Delta=|\Delta| e^{i\theta}$ is the order parameter. The annihilation operators in the band basis $(\psi_{ck}, \psi_{vk})$, the `paramagnetic current' $\vec{j}_k$, the renormalized quasi-particle energy $E_k$, and the bare velocity $\vec{v}$ are
\begin{align}
	&
	\begin{pmatrix}
		\psi_{ck}
		\\
		\psi_{vk}
	\end{pmatrix}
	=
	\begin{pmatrix}
		u_k^\ast & v_k^\ast
		\\
		-v_k & u_k
	\end{pmatrix}
	\begin{pmatrix}
		\psi_{1k}
		\\
		\psi_{2k}
	\end{pmatrix}
	\,,\quad
	(u_k,\, v_k)= 
	\frac{1}{\sqrt{2}}\left(
	\sqrt{1+\frac{\xi_k}{E_k}},\, e^{i\theta} \sqrt{1-\frac{\xi_k}{E_k}}
	\right)
	\notag\\
	&	\vec{j}_k=
	\frac{\vec{v}_k}{E_k}
	\begin{pmatrix}
		\xi_k & \Delta \\
		\Delta^\ast  &  -\xi_k
	\end{pmatrix}
	\,,\quad
	E_k=\sqrt{\xi_k^2+ \Delta^2}
	\,,\quad
	\vec{v}_k =\partial_{\vec{k}} \xi_k
	\,.
	\label{eqnSI:BCS_HS_2}
\end{align}

To obtain the ponderomotive force $F_{\text{P}}$ for the order parameter  $\Delta$,  we integrate out the fermions $\psi$ in the Keldysh path integral and look for $O(\Delta_q)$ terms in the resulting action, as indicated by \equa{eqnSI:action_phi}. Without losing important information, we focus on the force on the amplitude direction and take $\Delta$ to be real. 
$\Delta$ enters the driven fermions in two places: 
the periodic driving term $\vec{A} \cdot \vec{j}$, and the quasi particle energy $E_k$. Taking a partial derivative with respect to $\Delta$, one obtains the force operator:
\begin{align}
	\hat{F}_P  = 	\hat{F}_1+\hat{F}_2
	\,,\quad
	\hat{F}_{1}= 
	\sum_k 
	\vec{A} \cdot \vec{v}_k
	\begin{pmatrix}
		\psi_{ck}^\dagger 
		&
		\psi_{vk}^\dagger 
	\end{pmatrix}
	\left[
	\partial_\Delta
	\begin{pmatrix}
		\frac{	\xi_k}{E_k} &  \frac{\Delta}{E_k}  \\
		\frac{\Delta}{E_k}  \ &  - \frac{	\xi_k}{E_k} 
	\end{pmatrix}
	\right]
	\begin{pmatrix}
		\psi_{ck}
		\\
		\psi_{vk}
	\end{pmatrix}
	\,,\quad
	\hat{F}_{2}= 
	\sum_k 
	\partial_\Delta  E_k
	\begin{pmatrix}
		\psi_{ck}^\dagger 
		&
		\psi_{vk}^\dagger 
	\end{pmatrix}
	\sigma_3
	\begin{pmatrix}
		\psi_{ck}
		\\
		\psi_{vk}
	\end{pmatrix}
	\,.
	\label{eqnSI:BCS_force}
\end{align}
The ponderomotive force $F_{\text{P}}=F_1+F_2$ is the path integral average of $\hat{F}_P $.

In this paper, we  compute $F_{\text{P}}$ only at order $A_0^2=E_0^2/\omega^2$. 
At this order, the $\hat{F}_{1}$ term contributes only by two-point correlators of $\vec{j}$ which has contributions only from the interband terms $\vec{v}_k  \frac{\Delta}{E_k} \left(\psi_{ck}^\dagger \psi_{vk} + c.c. \right)$.
The corresponding ponderomotive force is  found simply from the retarded two-point correlation function:
\begin{align}
	F_1&=
	A_0^2 \int dt  e^{i\omega t} \langle 
	j_{q} (t) F_{1c}(0)
	\rangle
	=
	A_0^2 \sum_k 
	v_{xk}^2
	\frac{\Delta}{E_k}
	\left( \partial_\Delta 
	\frac{\Delta}{E_k}
	\right)
	\sum_\Omega
	\mathrm{Tr} \left[
	\hat{G}_k(\Omega) \hat{\gamma}_1  \hat{\gamma}_c
	\hat{G}_k(\Omega+\omega) \hat{\gamma}_j  \hat{\gamma}_q
	\right]
	\notag \\
	&=
	A_0^2 \sum_k 
	v_{xk}^2
	\frac{\Delta}{E_k}
	\left( \partial_\Delta 
	\frac{\Delta}{E_k}
	\right)
	\mathrm{Re}
	\left[
	\frac{
		-8 E_k
	}
	{(\omega+i\gamma)^2 - 4E_k^2}
	\right]
	=
	-A_0^2
	\frac{v_F^2}{d} \nu \frac{4}{\Delta}
	\int d\xi 
	\frac{2\Delta^2 \xi^2}{E^3}
	\mathrm{Re}
	\left[
	\frac{1
	}
	{(\omega+i\gamma)^2 - 4E^2}
	\right]
	\notag\\
	&
	\equiv 
	-A_0^2
	\frac{v_F^2}{d}  \frac{4}{\Delta} \nu
	f_1\left(\frac{\Delta}{\omega},\, \frac{\gamma}{\omega}\right)
	\label{eqnSI:F2}
\end{align}
where $F_{1c}$ and $j_q$ are the `classical' and `quantum' components~\cite{Kamenev.book,Altland.2010} of these operators defined in the same way as those below \equa{eqnSI:Z},
$E=\sqrt{\Delta^2 + \xi^2}$, $\gamma$ is the quasi particle damping rate, 
$d$ is the space dimension,
and $\nu$ is the density of states at the fermi energy before the gap opening. The $\hat{\gamma}_c=\tau_0 
\sigma_0$,  $\hat{\gamma}_q=\tau_1 \sigma_0$,
$\hat{\gamma}_j=\tau_0 \sigma_1$,
$\hat{\gamma}_1=\tau_0 \sigma_1$ are the vertices for the classical source, quantum source, the current, and the force $F_1$.
The $4 \times 4$ Green's function 
\begin{align}
	\hat{G}_k(\omega)= 
	-i \langle 
	\psi \bar{\psi}
	\rangle|_\omega
	=
	\begin{pmatrix}
		G_k^R(\omega) & G_k^K(\omega) 
		\\
		0 & G_k^A(\omega) 
	\end{pmatrix}
	= 
	\begin{pmatrix}
		(\omega+i\gamma-E_k \sigma_3 )^{-1} & 
		\left( G_k^R(\omega) -G_k^A(\omega)  \right) 
		\tanh \frac{\omega}{2T} 
		\\
		0 & 	(\omega-i\gamma-E_k \sigma_3 )^{-1} 
	\end{pmatrix}
\end{align}
is in the Keldysh notation and in the band basis.

The $\hat{F}_{2}$ term contributes to the force by  three-point correlators:
\begin{align}
	F_2&=
	A_0^2 \int dt_1 dt_2  e^{i\omega (t_1-t_2)}  
	\langle 
	j_{q}(t_1) 	j_{q}(t_2)  F_{2c}(0)
	\rangle
	=
	A_0^2 
	\sum_k 
	\left(v_{xk}  \frac{\Delta}{E_k}\right)^2  
	\left(\partial_\Delta  E_k \right)
	\sum_\Omega
	\mathrm{Tr} \left[
	\hat{G}(\Omega) \hat{\gamma}_c  \hat{\gamma}_j
	\hat{G}(\Omega+\omega) \hat{\gamma}_c \hat{\gamma}_j
	\hat{G}(\Omega) \hat{\gamma}_q \hat{\gamma}_2
	\right]
	\notag \\
	&=
	A_0^2 
	\sum_k 
	\left(v_{xk}  \frac{\Delta}{E_k}\right)^2  
	\left(\partial_\Delta  E_k \right)
	\sum_\Omega
	\mathrm{Tr}
	\left[
	G_k^R(\Omega) G_k^K(\Omega) G_k^R(\Omega+\omega)
	+ 
	G_k^R(\Omega) G_k^A(\Omega) G_k^K(\Omega+\omega)
	+ 
	G_k^K(\Omega) G_k^A(\Omega) G_k^A(\Omega+\omega)
	\right]
	\notag \\
	&=
	A_0^2 
	\sum_k 
	\left(v_{xk}  \frac{\Delta}{E_k}\right)^2  
	\left(\partial_\Delta  E_k \right)
	\frac{2}{3}
	\left(
	\frac{
		2\tanh\left(\frac{-E_k}{T}\right)
		-\tanh\left(\frac{E_k}{T}\right) 
	}{(\omega-2E_k)^2 + \gamma^2}
	+ \frac{
		\tanh\left(\frac{-E_k}{T}\right)
		-2\tanh\left(\frac{E_k}{T}\right) 
	}{(\omega+2E_k)^2 + \gamma^2}
	\right)
	\notag\\
	&\xrightarrow{T=0}
	A_0^2 
	\sum_k 
	\left(v_{xk}  \frac{\Delta}{E_k}\right)^2  
	\left(\partial_\Delta  E_k \right)
	\left[
	-2 \left(
	\frac{1}{(\omega-2E_k)^2 + \gamma^2}
	+
	\frac{1}{(\omega+2E_k)^2 + \gamma^2}
	\right)
	\right]
	\notag\\
	&=
	-A_0^2
	\frac{v_F^2}{d}  \frac{4}{\Delta} \nu
	\int d\xi \frac{\Delta^4}{2E^3}
	\left(
	\frac{1}{(\omega-2E)^2 + \gamma^2}
	+ \frac{1}{(\omega+2E)^2 + \gamma^2}
	\right)
	\equiv
	-A_0^2
	\frac{v_F^2}{d}  \frac{4}{\Delta} \nu
	f_2\left(\frac{\Delta}{\omega},\, \frac{\gamma}{\omega}\right)
	\label{eqnSI:F1}
\end{align}
where $\hat{\gamma}_2=\tau_0 \sigma_3$.  

From \equa{eqnSI:F1}, it is seen that each quasi particle excitation (may also be viewed as the resonant excitation of a two level system) contributes a pole  $1/((\omega-2E)^2 + \gamma^2)$ to $F_{\text{P}}$ whose spectra weight scales as $1/\gamma$. This is a canonical example of the dissipative contribution to the  ponderomotive force when the slow field enters by shifting the energy $2E$ of a two-level system.

The total ponderomotive force is  therefore
\begin{align}
	F_{\text{P}}&=
	A_0^2 \sum_k v_{xk}^2
	\Bigg\{
	\frac{\Delta}{E_k}
	\left( \partial_\Delta 
	\frac{\Delta}{E_k}
	\right)
	\mathrm{Re}
	\left[
	\frac{-8E_k}{(\omega+i\gamma)^2 - 4E_k^2}
	\right]
	+2
	\frac{\Delta^2 \partial_\Delta E_k}{E_k^2} 
	\left[
	\frac{-1}{(\omega-2E_k)^2+\gamma^2}
	+ (E_k \rightarrow -E_k)
	\right]
	\Bigg\}
	\notag\\
	&=F_1+F_2=
	-A_0^2
	\frac{v_F^2}{d}   \nu \frac{4}{\Delta}
	\left[
	f_1\left(\frac{\Delta}{\omega},\, \frac{\gamma}{\omega}\right)
	+
	f_2\left(\frac{\Delta}{\omega},\, \frac{\gamma}{\omega}\right)
	\right]
	\,
	\label{eqnSI:BCS_F_P}
\end{align}
and  the  ponderomotive potential is
\begin{align}
	V_{\text{P}}= - \int F_{\text{P}} d\Delta
	=
	V_u \left(E_0, \omega \right)
	g_P\left(\frac{\Delta}{\omega},\, \frac{\gamma}{\omega}\right)
	\,,\quad
	V_u= 	
	\frac{ \nu \omega^2}{d}\left(
	\frac{e E_0}{\omega^2/v_F}
	\right)^2
	,\quad
	g_P\left(t,\, \delta \right)
	=
	\int_\infty^t dx \frac{4}{x}
	\left[
	f_1\left(x,\, \delta\right)
	+
	f_2\left(x,\, \delta\right)
	\right]
	\,
	\label{eqnSI:BCS_V_P}
\end{align}
where $V_u$ is a scale of energy density determined by the driving field $E_0$ and frequency $\omega$, and $g_P$, $f_1$, $f_2$ are dimensionless functions of $\Delta/\omega$ and $\gamma/\omega$ only.

In \fig{figSI:BCS}(d), we plot the dimensionless  $F_{\text{P}}$ and $V_{\text{P}}$. The negative sign of $F_{\text{P}}$ means that it always tends to push $\Delta$ to smaller values. As $\Delta$ decreases to around $\omega/2$,  there is a strong peak in $F_{\text{P}}$ due to the quasi-particle excitations (see also the optical conductivity in \fig{figSI:BCS}(a)), which results in a sharp drop of $V_{\text{P}}$ to its value $\sim -1/\gamma$ at $\Delta=0$. This combined with the equilibrium contribution to the free energy $F_0$ results in the new free energy minimum in $F_0+V_{\text{P}}$, see \fig{fig:landscape}. We also show the asymptotic values of $F_{\text{P}}$ and $V_{\text{P}}$ in different regimes of \fig{figSI:BCS}, which gives the schematic phase diagram in the inset of \fig{fig:landscape}.

We note that in principle, in the resonance regime $2\Delta<\omega$, the expansion of $V_{\text{P}}$ at the order of $E_0^2$ is a good approximation  when $\kappa = e E_0 v_F/(\omega \gamma) \ll 1$. To obtain the corrections at large fields, one may employ the exact solutions for the driven-dissipative two levels system, one for each Anderson-pseudo spin. We leave it for future study.

\subsection{Insights from the optical conductivity}
One may check that in the dissipationless limit, meaning $\gamma \rightarrow 0$ and $\omega < 2\Delta$, the ponderomotive potential (\equa{eqnSI:BCS_F_P}) reduces to that predicted by Lemma~1:
\begin{align}
	V_{\text{P}}= 
	\mathrm{Re}
	\left[-\frac{i}{\omega} \sigma(\Delta,\omega)
	\right] 
	E_0^2
	\,.
	\label{eqnSI:F_P_no_gamma}
\end{align}
where $\sigma$ may be found from Eq.~12 of Ref.~\cite{sun.2020_BaSh}, also shown in \fig{figSI:BCS}(a).
According to \equa{eqnSI:F_P_no_gamma}, the behavior of $V_{\text{P}}$ at $\Delta > \omega/2$ (black curve in \fig{figSI:BCS}(d)) may be understood from the sub-gap optical conductivity (blue curve at $\omega < 2\Delta$ in \fig{figSI:BCS}(a)). Giving a driving frequency $\omega < 2\Delta$,  it is obvious that if the gap  decreases, the $V_{\text{P}}$ in  \equa{eqnSI:F_P_no_gamma} would drop. 

\equa{eqnSI:F_P_no_gamma} also explains the light tuned competing superconducting and charge orders in the attractive Hubbard model discussed by Sentef et al.~\cite{Sentef.2017_SC_CDW}.
If the spectra weight shift due to the phase mode is ignored as in Ref.~\cite{Sentef.2017_SC_CDW}, the CDW state has an optical conductivity shown in \fig{figSI:BCS}(a), while the superconducting state has no optical absorptions ($\sigma=ine^2/(m \omega)$).
When driven with a sub-gap frequency $\omega < 2\Delta$, the CDW state gains a negative $V_{\text{P}}$  while the superconducting state has a positive $V_{\text{P}}$, as shown by \equa{eqnSI:F_P_no_gamma}. Therefore, the drive favors the CDW state.
When driven with a frequency right above the gap, assuming \equa{eqnSI:F_P_no_gamma} still works, the $V_{\text{P}}$ of the CDW state jumps to the a positive value, and may instead favor the superconducting state.

\end{widetext}

\bibliography{references/Nonequilibrium.bib,
references/My_publication.bib, references/Excitonic_Insulator.bib,
references/Excitons.bib, references/Excitonic_Insulator_2.bib,
references/Cavity.bib,
references/Parametric_Amplification.bib,
references/Superconductivity.bib,
references/RG.bib,
references/CDW.bib,
references/Ponderomotive_Force.bib,
references/notes.bib,
references/DiracFluid_library.bib,
references/Nonequilibrium2.bib}

\end{document}